\documentclass[journal]{IEEEtran}
\usepackage[noadjust]{cite}

\ifCLASSINFOpdf
\else
\fi
\usepackage{amsmath}
\usepackage{amssymb}
\usepackage{mathtools}
\usepackage{accents}
\usepackage{verbatim}
\usepackage{enumitem,kantlipsum}
\usepackage{color}

\usepackage[caption=false,font=footnotesize]{subfig}

\newtheorem{prob}{Problem}
\newtheorem{lemma}{Lemma}
\newtheorem{definition}{Definition}

\newtheorem{example}{Example}
\newtheorem{remark}{Remark}
\newtheorem{theorem}{Theorem}

\newenvironment{Note to Practitioners}
{\begin{abstract}}
	{\end{abstract}}

\begin{document}
%
\title{Motion and Cooperative Transportation Planning for Multi-Agent Systems under Temporal Logic Formulas}
%
%
%

\author{Christos K. Verginis,~\IEEEmembership{Member,~IEEE,}        
        and ~Dimos V. Dimarogonas,~\IEEEmembership{Member,~IEEE}
\thanks{The authors are is with the ACCESS Linnaeus Center, School of Electrical Engineering,
KTH Royal Institute of Technology, SE-100 44, Stockholm, Sweden and
with the KTH Center for Autonomous Systems. Email: {cverginis,
anikou, dimos}@kth.se.  This  work  was  supported  by  the
H2020  ERC  Starting  Grant  BUCOPHSYS,  the  Swedish  Research  Council
(VR),  the  Knut  och  Alice  Wallenberg  Foundation,  the  EU
H2020 Research and Innovation Programme under the GA  No.  644128  (AEROWORKS)  and  the  EU  H2020  Research  and
Innovation Programme under the GA No. 731869 (Co4Robots).}}
\maketitle

\begin{abstract}
This paper presents a hybrid control framework for the motion planning of a multi-agent system including $N$ robotic agents and $M$ objects, under high level goals expressed as Linear Temporal Logic (LTL) formulas. In particular, we design control protocols that allow the transition of the agents as well as the cooperative transportation of the objects by the agents, among predefined regions of interest in the workspace. This allows to abstract the coupled behavior of the agents and the objects as a finite transition system and to design a high-level multi-agent plan that satisfies the agents' and the objects' specifications, given as temporal logic formulas. Simulation results verify the proposed framework. 
\end{abstract}

\begin{Note to Practitioners}
	This paper is mainly motivated by scenarios that include multiple robots and objects of interest that have to be transported and/or processed in a certain way (e.g., manufacturing, rescue missions) according to temporal logic formulas. In contrast to existing methodologies, we define such objectives not only for the robots, but also for the objects (e.g., ``take object 1 to region A infintely many times while avoiding region B"). The key idea of our methodology lies in the construction of a discrete transition system that captures the motion of the agents and the objects around the workspace, which is based on the design of continuous control laws for agent navigation and cooperative object transportation. With this abstraction in hand, we are able to derive hybrid control protocols that satisfy the given temporal logic formulas for the coupled multi-agent system.
\end{Note to Practitioners}



%
\IEEEpeerreviewmaketitle

\section{Introduction}
%
%
%
%
\IEEEPARstart{T}emporal-logic based motion planning has gained significant amount of attention over the last decade, as it provides a fully automated  correct-by-design controller synthesis approach for autonomous robots. Temporal logics, such as linear temporal logic (LTL), provide formal high-level languages that can describe planning objectives more complex than the well-studied navigation algorithms, and have been used extensively both in single- as well as in multi-agent setups (see, indicatively, \cite{Fainekos2009,Lahijanian2016,Loizou2004,Diaz2015,Chen2012,Cowlagi2016,Belta2005,Bhatia2011,Filippidis2012,Meng15}). The objectives are given as a temporal logic formula with respect to a discretized abstraction of the system (usually a finite transition system), and then, a high-level discrete path is found by off-the-shelf model-checking algorithms, given the abstracted system and the task specification \cite{baier2008principles}.  

Most works in the related literature consider temporal logic-based motion planning for fully actuated, autonomous agents. Consider, however, cases where some unactuated objects must undergo a series of processes in a workspace with autonomous agents (e.g., car factories). In such cases, the agents, except for satisfying their own motion specifications, are also responsible for coordinating with each other in order to transport the objects around the workspace. When the unactuated objects' specifications are expressed using temporal logics, then the abstraction of the agents' behavior becomes much more complex, since it has to take into account the objects' goals. 

In addition, the spatial discretization of a multi-agent system to an abstracted higher level system necessitates the design of appropriate continuous-time controllers for the transition of the agents among the states of discrete system. Many works in the related literature, however, either assume that there \textit{exist} such continuous controllers or adopt non-realistic assumptions. For instance, many works either do not take into account continuous agent dynamics or consider single or double integrators \cite{Loizou2004,Filippidis2012,Fainekos2009,Bhatia2011,Meng15}, which can deviate from the actual dynamics of the agents, leading thus to poor performance in real-life scenarios. 
Discretized abstractions, including design of the discrete state space and/or continuous-time controllers, can be found in \cite{Belta2005,belta2006controlling,reissig2011computing,tiwari2008abstractions,rungger2015state} for general systems and \cite{boskos2015decentralized,belta2004abstraction} for multi-agent systems. Moreover, many works adopt dimensionless point-mass agents and therefore do not consider inter-agent collision avoidance \cite{Belta2005,Filippidis2012,Meng15}, which can be a crucial safety issue in applications involving autonomous robots. 

Since we aim at incorporating the unactuated objects' specifications in our framework, the agents have to perform (cooperative) transportation of the objects around the workspace, while avoiding collisions with each other. Cooperative transportation/manipulation has been extensively studied in the literature (see, for instance, \cite{sugar2002control,heck2013internal,kume2007coordinated,tsiamis2015cooperative,ficuciello2014cartesian,ponce2016cooperative,marino2017distributed,nikou2017nonlinear,erhart2016model}), with collision avoidance specifications being incorporated in \cite{tanner2003nonholonomic}, which is the main inspiration of our cooperative transportation methodology. Cooperative object transportation under temporal logics has also been considered in our previous work \cite{verginis2017distributed}. 

This paper presents a novel hybrid control framework for the motion planning of a team of $N$ autonomous agents and $M$ unactuated objects under LTL specifications. Using previous results on navigation functions, we design feedback control laws for i) the navigation of the agents and ii) the cooperative transportation of the objects by the agents, among predefined regions of interest in the workspace, while ensuring inter-agent collision avoidance. This allows us to model the coupled behavior of the agents and the objects with a finite transition system, which can be used for the design of high-level plans that satisfy the given LTL specifications.  This paper is an extension of our previous work \cite{verginis2017ifacPlann}, where we did not account for cooperative transportation of the objects. 

The rest of the paper is organized as follows. Section \ref{sec:Notation-and-Preliminaries} provides necessary notation and preliminary background. The problem is formulated in Section \ref{sec:Model and PF} and the proposed solution is presented in Section \ref{sec:main results}. Finally, Section \ref{sec:simulations} provides simulation results and Section \ref{sec:conclusion} concludes the paper.

\section{Notation and Preliminaries} \label{sec:Notation-and-Preliminaries}
\subsection{Notation} \label{subsec:Notation}
Vectors and matrices are denoted with bold lowercase and uppercase letters, respectively, whereas scalars are denoted with non-bold lowercase letters.
The set of positive integers is denoted as $\mathbb{N}$ and the real $n$-space, with $n\in\mathbb{N}$, as $\mathbb{R}^n$;
$\mathbb{R}^n_{\geq 0}$ and $\mathbb{R}^n_{> 0}$ are the sets of real $n$-vectors with all elements nonnegative and positive, respectively. We also use $\mathbb{T} = (-\pi,\pi)\times(-\tfrac{\pi}{2},\tfrac{\pi}{2})\times(-\pi,\pi)$.  Given a set $S$, $\mathring{S}$ is its interior, $2^S$ is the set of all possible subsets of $S$, $\lvert S \rvert$ is its cardinality, and, given a finite sequence $s_1,\dots,s_n$ of elements in $S$, with $n\in\mathbb{N}$, we denote by $(s_1,\dots,s_n)^\omega$ the infinite sequence $s_1,\dots,s_n,s_1,\dots,s_n,s_1,\dots s_n,\dots$ created by repeating $s_1,\dots,s_n$. The notation $\|\boldsymbol{y}\|$ is used for the Euclidean norm of a vector $\boldsymbol{y} \in \mathbb{R}^n$; $SO(3)$ is the $3$D rotation group and $\boldsymbol{S}:\mathbb{R}^3\to\mathbb{R}^{3\times3}$ is the skew-symmetric matrix derived by the relation $\boldsymbol{S}(\boldsymbol{x})\boldsymbol{y} \coloneqq \boldsymbol{x}\times\boldsymbol{y}$, where $\times$ is the cross-product operator. Given a nonempty and bounded set of natural numbers $\mathcal{X}$ and a set of vectors (matrices) $\boldsymbol{x}_i, i\in\mathbb{N}$, we denote by $[\boldsymbol{x}^\top_i]^\top_{i\in\mathcal{X}}$  the stack column-vector form with the vectors (matrices) whose indices belong to $\mathcal{X}$. 
Given $x\in\mathbb{R}$ and $\boldsymbol{y},\boldsymbol{z}\in\mathbb{R}^n$, we use $\nabla_{\boldsymbol{z}}x \coloneqq \partial x/\partial \boldsymbol{z}\in\mathbb{R}^n$ and  $\nabla_{\boldsymbol{z}}\boldsymbol{y} \coloneqq \partial \boldsymbol{y}/\partial \boldsymbol{z}\in\mathbb{R}^{n\times n}$;
$\mathcal{B}_n:\mathbb{R}^{n}\times\mathbb{R}_{\geq 0}\rightrightarrows \mathbb{R}^n$ is the set-valued map that represents the closed ball $\mathcal{B}(\boldsymbol{c},r) \coloneqq \{x\in\mathbb{R}^n : \|\boldsymbol{x} - \boldsymbol{c}\| \leq r\}$ of center $\boldsymbol{c}$ and radius $r$. In addition, we use $\mathcal{N} \coloneqq \{1,\dots,N\},\mathcal{M} \coloneqq \{1,\dots,M\},\mathcal{K} \coloneqq \{1,\dots,K\}$, with $N,M,K\in\mathbb{N}$, as well as $\mathbb{M}\coloneqq\mathbb{R}^3\times\mathbb{T}$. Finally, all differentiations are expressed with respect to an inertial reference frame $\{I\}$, unless otherwise stated.  
\subsection{Task Specification in LTL} \label{subsec:LTL}

We focus on the task specification $\phi$ given as a Linear Temporal Logic (LTL) formula. The basic ingredients of a LTL formula are a set of atomic propositions $\Psi$ and several boolean and temporal operators. LTL formulas are formed according to the following grammar \cite{baier2008principles}: $\phi ::= \mathsf{true}\: |\:a\: |\: \phi_{1} \land  \phi_{2}\: |\: \neg \phi\: |\:\bigcirc \phi\:|\:\phi_{1}\cup\phi_{2} $, where $a\in \Psi$, $\phi_1$ and $\phi_2$ are LTL formulas and $\bigcirc$, $\cup$ are the \textit{next} and \textit{until} operators, respectively. Definitions of other useful operators like $\square$ (\it always\rm), $\lozenge$ (\it eventually\rm) and $\Rightarrow$ (\it implication\rm) are omitted and can be found at \cite{baier2008principles}.
The semantics of LTL are defined over infinite words over $2^{\Psi}$. Intuitively, an atomic proposition $\psi\in \Psi$ is satisfied on a word $w=w_1w_2\dots$ if it holds at its first position $w_1$, i.e. $\psi\in w_1$. Formula $\bigcirc\phi$ holds true if $\phi$ is satisfied on the word suffix that begins in the next position $w_2$, whereas $\phi_1\cup\phi_2$ states that $\phi_1$ has to be true until $\phi_2$ becomes true. Finally, $\lozenge\phi$ and  $\square\phi$ holds on $w$ eventually and always, respectively. For a full definition of the LTL semantics, the reader is referred to \cite{baier2008principles}.

\subsection{Multirobot Navigation Functions (MRNFs)} \label{subsec:MAS NF}
Navigation functions, initially proposed in \cite{Koditchek92} for single-point-sized robot navigation, are real-valued maps realized through cost functions, whose negated gradient field is attractive towards the goal configuration (referred to as the good or desirable set) and repulsive with respect to the obstacles set (referred to as the bad set which we want to avoid). Multirobot Navigation Functions (MRNFs) were developed in \cite{Loizou2006}, for which we provide here a brief overview.

Consider $N\in\mathbb{N}$ spherical robots, with center $\boldsymbol{q}_i\in\mathbb{R}^n$, $n\in\mathbb{N}$, and radius $r_i \in\mathbb{R}_{>0}$, i.e., $\mathcal{B}_n(\boldsymbol{q}_i,r_i)$, $i\in\mathcal{N}$, operating in an open spherical workspace $\mathcal{W}\coloneqq \mathring{\mathcal{B}}_n(\boldsymbol{0},r_0)$ of radius $r_0 \in\mathbb{R}_{>0}$. Each robot has a destination point $\boldsymbol{q}_{\text{d}_i}\in\mathbb{R}^n, i\in\mathcal{N}$, and $\boldsymbol{q}_{\text{d}} \coloneqq [\boldsymbol{q}^\top_{\text{d}_1},\dots, \boldsymbol{q}^\top_{\text{d}_N}]^\top$.
Let $\mathcal{F}\subset \mathbb{R}^n$ be a compact connected analytic manifold with boundary. A map $\varphi: \mathcal{F}\to[0,1]$ is a \textit{MRNF} if 
\begin{enumerate}
\item It is analytic on $\mathcal{F}$,
\item It has only one minimum at $\boldsymbol{q}_\text{d}\in \overset{\circ}{F}$,
\item Its Hessian at all critical points is full rank,
\item $\lim\limits_{\boldsymbol{q}\to \partial \mathcal{F}} = 1 > \varphi(\boldsymbol{q}')$, $\forall \boldsymbol{q}'\in \overset{\circ}{F}$,
\end{enumerate}
where $\boldsymbol{q}\coloneqq [\boldsymbol{q}^\top_1,\dots,\boldsymbol{q}^\top_N]^\top\in\mathbb{R}^{Nn}$. The class of MRNFs has the form
\small
\begin{equation*}
\varphi(\boldsymbol{q}) = \frac{\gamma(\boldsymbol{q})}{\Big( [\gamma(\boldsymbol{q})]^{\kappa} + G(\boldsymbol{q}) \Big)^{\tfrac{1}{\kappa}}},
\end{equation*}
\normalsize
where $\gamma(\boldsymbol{q}) \coloneqq \| \boldsymbol{q} - \boldsymbol{q}_{\text{d}} \|^2$ is the goal function, $G(\boldsymbol{q})$ is the obstacle function, and $\kappa$ is a tunable gain; $\gamma^{-1}(0)$ denotes the desirable set and $G^{-1}(0)$ the set we want to avoid. Next we provide the procedure for the construction of the function $G$. A robot proximity function, a measure for the distance between two robots $i,l\in\mathcal{N}$, is defined as $\beta_{i,l}(\boldsymbol{q}_i,\boldsymbol{q}_l) \coloneqq \|\boldsymbol{q}_i - \boldsymbol{q}_l\|^2 - (r_i + r_l)^2$, $\forall i,l\in\mathcal{N}, i\neq l$. The term \textit{relation} is used to describe the possible collision schemes that can be defined in a multirobot team, possibly including obstacles.
The \textit{set of relations} between the members of the team can be defined as the set of all possible collision schemes between the members of the team. A binary relation is a relation between two robots. Any relation can be expressed as a set of binary relations. A \textit{relation tree} is the set of robot/obstacles that form a linked team. Each relation may consist of more than one relation tree. The number of binary  relations in  a  relation is called \textit{relation level}. Illustrative examples can be found in \cite{Loizou2006}. A \textit{relation proximity function} (RPF) provides a measure of the distance between the robots involved in a relation. Each relation has its own RPF. A RPF is the sum of the robot proximity functions of a relation.
It assumes the value of zero whenever the related robots collide (since the involved robot proximity functions will be zero) and increases with respect to the distance of the related robots. The RPF of relation $j$ at level $k$ is given by $(b_{R_j})_k \coloneqq \sum\limits_{(i,m)\in(R_j)_k}\beta_{i,m}$, where we omit the arguments $\boldsymbol{q}_i,\boldsymbol{q}_k$ for notational brevity.  A \textit{relation verification function} (RVF) is defined as 
\small
\begin{equation*}
g_{R_j} \coloneqq (b_{R_j})_k + \lambda \frac{ (b_{R_j})_k}{(b_{R_j})_k + ( B_{(R^C_j)_k} )^{\tfrac{1}{h}}}, 
\end{equation*}
\normalsize
where $\lambda, h > 0$, and $R^C_j$ is the complementary to $R_j$ set of relations in the same level $k$, $j$ is an index number defining the relation in level $k$, and $B_{R^C_j} \coloneqq \prod\limits_{m\in R^C_j} b_m$. The RVF serves as an analytic switch, which goes to zero only when the relation it represents is realized. By further introducing the workspace boundary obstacle functions as $G_0 \coloneqq \prod_{i\in\mathcal{N}}\Big\{ (r_0 -r_i)^2 - \|\boldsymbol{q}_i\|^2 \Big\}$, we can define 
$G \coloneqq G_0\prod_{L=1}^{n_L}\prod_{j=1}^{n_{R,L}}(g_{R_j})_L,$
where $n_L$ is the number of levels and $n_{R,L}$ the number of relations in level $L$. It has been proved that, by choosing the parameter $\kappa$ large enough, the negated gradient field $-\nabla_{\boldsymbol{q}}\varphi(\boldsymbol{q})$ leads to the destination configuration $\boldsymbol{q}_{\text{d}}$, from almost all initial conditions \cite{Loizou2006}.

\section{System Model and Problem Formulation} \label{sec:Model and PF}
Consider $N>1$ robotic agents operating in a workspace $\mathcal{W}$ with $M>0$ objects; $\mathcal{W}$ is a bounded open ball in $3$D space, i.e., $\mathcal{W}\coloneqq\mathring{\mathcal{B}}(\boldsymbol{0},r_0)=\{\boldsymbol{p}\in \mathbb{R}^3 \text{ s.t. } \lVert \boldsymbol{p} \rVert < r_0 \}$, where $r_0\in\mathbb{R}_{>0}$ is the radius  of $\mathcal{W}$. The objects are represented by rigid bodies whereas the robotic agents are fully actuated and consist of a fully actuated holonomic moving part (i.e., mobile base) and a robotic arm, having, therefore, access to the entire workspace.  Within $\mathcal{W}$ there exist $K>1$ smaller spheres around points of interest, which are described by $\mathcal{\pi}_k \coloneqq\mathcal{B}(\boldsymbol{p}_{\pi_k},r_{\pi_k}) = \{\boldsymbol{p}\in \mathbb{R}^3 \text{ s.t. } \lVert \boldsymbol{p}-\boldsymbol{p}_{\pi_k}\rVert\leq r_{\pi_k} \}$, where $\boldsymbol{p}_{\pi_k}\in \mathbb{R}^3$ is the center and $r_{\pi_k} \in\mathbb{R}_{>0}$ the radius of $\pi_k$. 
We denote the set of all $\pi_k$ as $\Pi \coloneqq \{\pi_1,\dots,\pi_K \}$. 
Moreover, we introduce disjoint sets of atomic propositions $\Psi_i, \Psi^{\scriptscriptstyle O}_j$, expressed as boolean variables, that represent services provided to agent $i\in\mathcal{N}$ and object $j\in\mathcal{M}$ in $\Pi$. The services provided at each region $\pi_k$ are given by the labeling functions $\mathcal{L}_i:\Pi\rightarrow2^{\Psi_i}, \mathcal{L}^{\scriptscriptstyle O}_j:\Pi\rightarrow2^{\Psi^{\scriptscriptstyle O}_j}$, which assign to each region $\pi_k, k\in\mathcal{K}$, the subset of services $\Psi_i$ and $\Psi^{\scriptscriptstyle O}_j$, respectively, that can be provided in that region to agent $i\in\mathcal{N}$ and object $j\in\mathcal{M}$, respectively. In addition, we consider that the agents and the object are initially ($t=0$) in the regions of interest $\pi_{init(i)}, \pi_{init_{\scriptscriptstyle O}(j)}$, where the functions $init:\mathcal{N}\to\mathcal{K}$, $init_{\scriptscriptstyle O}:\mathcal{M}\to\mathcal{K}$ specify the initial region indices. 
We denote by $\left\{ E_{i}\right\}$, $\left\{ O\right\}$
the robotic arms' end-effector and
object's center of mass frames, respectively; $\left\{ I\right\} $
corresponds to an inertial frame of reference. In the following, we present the modeling of the coupled kinematics and dynamics of the object and the agents.
%


We denote by $\boldsymbol{q}_i,\dot{\boldsymbol{q}}_i \in\mathbb{R}^{n_i}$, with $n_i\in\mathbb{N}, \forall i\in\mathcal{N}$, the generalized joint-space variables and their time derivatives for agent $i$. The overall joint configuration is then $\boldsymbol{q} \coloneqq [\boldsymbol{q}^\top_1,\dots,\boldsymbol{q}^\top _N]^\top$, $\dot{\boldsymbol{q}} \coloneqq [\dot{\boldsymbol{q}}^\top _1,\dots,\dot{\boldsymbol{q}}^\top _N]^\top \in\mathbb{R}^{n}$, with $n \coloneqq \sum_{i\in\mathcal{N}}n_i$. In addition, the inertial position and Euler-angle orientation of the $i$th end-effector, denoted by $\boldsymbol{p}_i$ and $\boldsymbol{\eta}_i$, respectively, expressed in an inertial reference frame, can be derived by the forward kinematics and are smooth functions of $\boldsymbol{q}_i$, i.e. $\boldsymbol{p}_i:\mathbb{R}^{n_i}\to\mathbb{R}^3$, $\boldsymbol{\eta}_i:\mathbb{R}^{n_i}\to\mathbb{T}$. The generalized velocity of each agent's end-effector $\boldsymbol{v}_i \coloneqq [\dot{\boldsymbol{p}}^\top_i,\boldsymbol{\omega}^\top_i]^\top \in\mathbb{R}^6$, can be considered as a transformed state through the differential kinematics $\boldsymbol{v}_i = \boldsymbol{J}_i(\boldsymbol{q}_i)\dot{\boldsymbol{q}}_i$ \cite{Siciliano2010}, where $\boldsymbol{J}_i:\mathbb{R}^{n_i}\to\mathbb{R}^{6\times n_i}$ is a smooth function representing the geometric Jacobian matrix, $\forall i\in\mathcal{N}$ \cite{Siciliano2010}. The matrix inverse of $\boldsymbol{J}_i$ is well defined in the set away from \textit{kinematic singularities} \cite{Siciliano2010}, which we define as $\mathbb{S}_i \coloneqq \{\boldsymbol{q}_i\in\mathbb{R}^{n_i}: \det(\boldsymbol{J}_i(\boldsymbol{q}_i)[\boldsymbol{J}_i(\boldsymbol{q}_i)]^\top) > 0\}$, $\forall i\in\mathcal{N}$.  

The differential equation describing the \textit{joint-space} and \textit{task-space} dynamics of each agent is \cite{Siciliano2010}:
\begin{subequations}\label{eq:manipulator dynamics}
\begin{align}
&\boldsymbol{B}_i(\boldsymbol{q}_i)\ddot{\boldsymbol{q}}_i+\boldsymbol{N}_i(\boldsymbol{q}_i,\dot{\boldsymbol{q}}_i)\boldsymbol{q}_i+ \boldsymbol{g_q}_i(\boldsymbol{q}_i) = \boldsymbol{\tau}_i-[\boldsymbol{J}_i(\boldsymbol{q}_i)]^\top\boldsymbol{f}_i,\label{eq:manipulator dynamics_joint_space} \\
&\boldsymbol{M}_i(\boldsymbol{q}_i)\dot{\boldsymbol{v}}_i+\boldsymbol{C}_i(\boldsymbol{q}_i,\dot{\boldsymbol{q}}_i)\boldsymbol{v}_i+ \boldsymbol{g}_i(\boldsymbol{q}_i) = \boldsymbol{u}_i-\boldsymbol{f}_i,\label{eq:manipulator dynamics_task_space}
\end{align}
\end{subequations}
where $\boldsymbol{B}_i:\mathbb{R}^{n_i}\to\mathbb{R}^{n_i\times n_i}$ is the positive definite inertia matrix, $\boldsymbol{N}_i:\mathbb{R}^{n_i}\times\mathbb{R}^{n_i}\to\mathbb{R}^{n_i\times n_i}$ is the Coriolis matrix, $\boldsymbol{g_q}_i:\mathbb{R}^{n_i}\to\mathbb{R}^{n_i}$ is the gravity term, and $\boldsymbol{f}_i\in\mathbb{R}^{6}$ is a vector of generalized forces in case of contact with the external environment. Definitions of the task space terms
$\boldsymbol{M}_i:\mathbb{S}_i\to\mathbb{R}^{6\times 6}$, $\boldsymbol{C}_i:\mathbb{S}_i\times\mathbb{R}^{n_i}\to\mathbb{R}^{6\times 6}$, $\boldsymbol{g}_i:\mathbb{S}_i\to\mathbb{R}^6$ can be found in \cite{Siciliano2010}. The task space wrench $\boldsymbol{u}_i$
is related to the joint torques $\boldsymbol{\tau}_i$ via $\boldsymbol{\tau}_i = \boldsymbol{J}^\top _i(\boldsymbol{q}_i)\boldsymbol{u}_i + (\boldsymbol{I}_{n_i} - \boldsymbol{J}^\top _i(\boldsymbol{q}_i)\widetilde{\boldsymbol{J}}^\top _i(\boldsymbol{q}_i))\boldsymbol{\tau}_{i0}$, $\forall i\in\mathcal{N}$; $\widetilde{\boldsymbol{J}}_i(\boldsymbol{q}_i) \coloneqq \boldsymbol{M}_i(\boldsymbol{q}_i)\boldsymbol{J}_i(\boldsymbol{q}_i)[\boldsymbol{B}_i(\boldsymbol{q}_i)]^{-1}$ is a generalized inverse and $\boldsymbol{\tau}_{i0}$ concerns redundant degrees of freedom ($n_i > 6$) and does not contribute to end-effector forces \cite{Siciliano2010}. The aforementioned dynamic terms are well-defined in the set $\mathbb{S}_i$, away from kinematic singularities. Avoidance of such configurations is not explicitly taken account in this work. Note, however, that the agents' tasks consist of navigating as well as cooperatively transporting the objects to predefined points in the workspace. This along with the fact that the agents consist of fully actuated moving bases imposes a kinematic redundancy, which can be exploited to avoid kinematic singularities,  (e.g., through the term $\boldsymbol{\tau}_{i0}$).  
 
We consider that each agent $i$, for a given $\boldsymbol{q}_i$, covers a spherical region $\mathcal{A}_i: \mathbb{R}^{n_i}\rightrightarrows \mathbb{R}^3$ of constant radius $r_i\in\mathbb{R}_{>0}$ that bounds its volume for that given $\boldsymbol{q}_i$, i.e., $\mathcal{A}_i(\boldsymbol{q}_i)\coloneqq \mathcal{B}(\boldsymbol{c}_i(\boldsymbol{q}_i),r_i)$, where $\boldsymbol{c}_i:\mathbb{R}^{n_i}\to\mathbb{R}^3$ is the center of the spherical region (a point on the robotic arm), $\forall i\in\mathcal{N}$; $\mathcal{A}_i$ can be obtained by considering the smallest sphere that covers the workspace of the robotic arm, extended with the mobile base part.
Moreover, we consider that the agents have specific power capabilities, which for simplicity, we match to positive integers $\zeta_i > 0$, $i\in\mathcal{N}$, via an analogous relation.

\begin{figure}
\centering
\includegraphics[scale = 0.25]{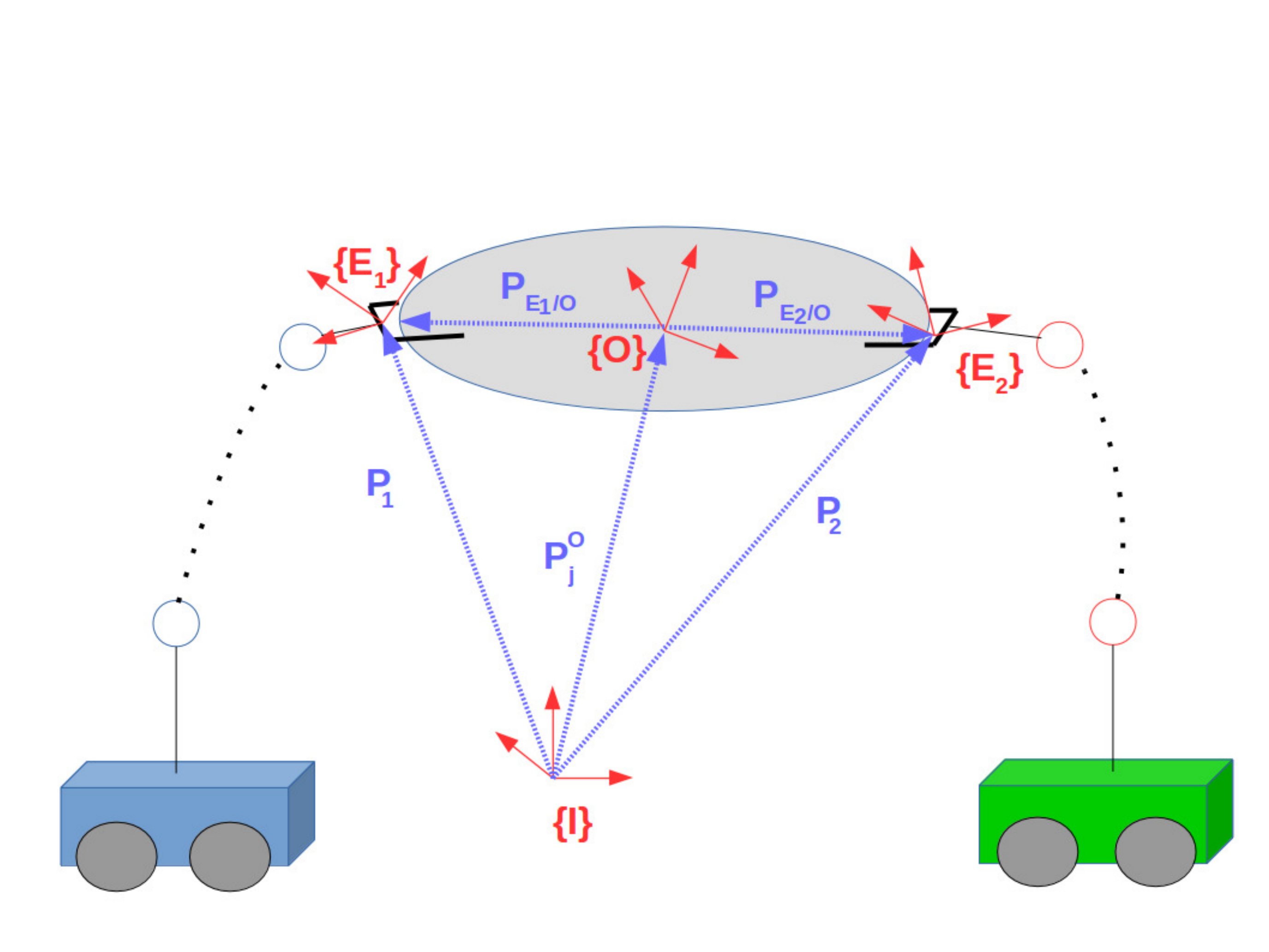}
\caption{Two agents rigidly grasping an object.  \label{fig:coop manip}}
\end{figure}


Regarding the objects, we denote by $\boldsymbol{x}^{\scriptscriptstyle O}_j\coloneqq [(\boldsymbol{p}^{\scriptscriptstyle O}_j)^\top,(\boldsymbol{\eta}^{\scriptscriptstyle O}_j)^\top]^\top \in\mathbb{M}$, $\boldsymbol{v}^{\scriptscriptstyle O}_j \coloneqq [(\dot{\boldsymbol{p}}^{\scriptscriptstyle O}_j)^\top, (\boldsymbol{\omega}^{\scriptscriptstyle O}_j)^\top]^\top \in\mathbb{R}^{12}$, $\forall j\in\mathcal{M}$, the pose (with $\boldsymbol{p}^{\scriptscriptstyle O}_j$ being the position of the center of mass with respect to (and expressed in) an inertial reference frame, and $\boldsymbol{\eta}^{\scriptscriptstyle O}_j\coloneqq[\eta^{\scriptscriptstyle O}_{j,1},\eta^{\scriptscriptstyle O}_{j,2},\eta^{\scriptscriptstyle O}_{j,3}]^\top$ denoting the extrinsic Euler angles) and generalized velocity of the $j$th object's center of mass, which is considered as the object's state. We consider the following second order dynamics, which can be derived based on the Newton-Euler formulation: 
\begin{subequations} \label{eq:object dynamics}
\begin{align}
& \dot{\boldsymbol{x}}^{\scriptscriptstyle O}_j = \boldsymbol{J}^{\scriptscriptstyle O}_j(\boldsymbol{x}^{\scriptscriptstyle O}_j)\boldsymbol{v}^{\scriptscriptstyle O}_j, \label{eq:object dynamics 1}\\
& \boldsymbol{M}_{\scriptscriptstyle O}(\boldsymbol{x}^{\scriptscriptstyle O}_j)\dot{\boldsymbol{v}}^{\scriptscriptstyle O}_j+\boldsymbol{C}_{\scriptscriptstyle O}(\boldsymbol{x}^{\scriptscriptstyle O}_j,\boldsymbol{v}^{\scriptscriptstyle O}_j)\boldsymbol{v}^{\scriptscriptstyle O}_j+\boldsymbol{g}_{\scriptscriptstyle O}(\boldsymbol{x}^{\scriptscriptstyle O}_j) = \boldsymbol{f}^{\scriptscriptstyle O}_j, \label{eq:object dynamics 2}
\end{align}
\end{subequations}
where $\boldsymbol{M}_{\scriptscriptstyle O}:\mathbb{M}\to\mathbb{R}^{6\times6}$ is the positive definite inertia matrix, $\boldsymbol{C}_{\scriptscriptstyle O}:\mathbb{M}\times\mathbb{R}^{6}\to\mathbb{R}^{6\times6}$ is the Coriolis matrix, $\boldsymbol{g}_{\scriptscriptstyle O}:\mathbb{M}\to\mathbb{R}^6$ is the gravity vector, and $\boldsymbol{f}^{\scriptscriptstyle O}_j\in\mathbb{R}^6$ is the vector of generalized forces in case of contact with the external environment, $\forall j\in\mathcal{M}$. Moreover, $\boldsymbol{J}^{\scriptscriptstyle O}_j(\boldsymbol{x}^{\scriptscriptstyle O}_j)$ is only defined in the subset of $\mathbb{M}$ that does not include the configurations where the pitch angle $\eta^{\scriptscriptstyle O}_{j,2}$ is $\pm \tfrac{\pi}{2}$, namely, \textit{representation singularities}, i.e., $\boldsymbol{J}^{\scriptscriptstyle O}_j:\mathbb{S}^{\scriptscriptstyle O}_j\to\mathbb{R}^{6\times6}$, with $\mathbb{S}^{\scriptscriptstyle O}_j \coloneqq \{\boldsymbol{x}^{\scriptscriptstyle O}_j\in\mathbb{M} : |\eta^{\scriptscriptstyle O}_{j,2} | < \tfrac{\pi}{2} \}$, $\forall j\in\mathcal{M}$.

Similarly to the agents, each object's volume is represented by the spherical set $\mathcal{O}_j: \mathbb{R}^3 \rightrightarrows \mathbb{R}^3$ of a constant radius $r^{\scriptscriptstyle O}_j \in\mathbb{R}_{>0}$, i.e., $\mathcal{O}_j(\boldsymbol{x}^{\scriptscriptstyle O}_j)  \coloneqq \mathcal{B}(\boldsymbol{x}^{\scriptscriptstyle O}_j,r^{\scriptscriptstyle O}_j)$, $\forall j\in\mathcal{M}$. 


Next, we provide the coupled dynamics between an object $j\in\mathcal{M}$ and a subset $\mathcal{T}\subseteq\mathcal{N}$ of agents that grasp it rigidly (see Fig. \ref{fig:coop manip}).
In view of Fig. \ref{fig:coop manip}, one concludes that the pose of the agents and the object's center of mass are related as  
\begin{subequations} \label{eq:coupled_kinematics}
\begin{align}
\boldsymbol{p}_i(\boldsymbol{q}_i) &= \boldsymbol{p}^{\scriptscriptstyle O}_j + \boldsymbol{R}_i(\boldsymbol{q}_i) \boldsymbol{p}^{\scriptscriptstyle E_i}_{\scriptscriptstyle E_{i}/O_j},\label{eq:coupled_kinematics_1} \\
\boldsymbol{\eta}_i(\boldsymbol{q}_i) &= \boldsymbol{\eta}^{\scriptscriptstyle O}_j + \boldsymbol{\eta}_{\scriptscriptstyle E_i/O_j}, \label{eq:coupled_kinematics_2}
\end{align}
\end{subequations}
$\forall i\in\mathcal{T}$, where $\boldsymbol{R}_i:\mathbb{R}^{n_i}\to SO(3)$ is the rotation matrix from $\{I\}$ to the $i$th agent's end-effector $\{E_i\}$, and $\boldsymbol{p}^{\scriptscriptstyle E_i}_{\scriptscriptstyle E_i/O_j}$, $\boldsymbol{\eta}_{\scriptscriptstyle E_i/O_j}$ are the \textit{constant} distance and orientation offset between $\{O\}$ and $\{E_i\}$, respectively. 
Following \eqref{eq:coupled_kinematics}, along with the
fact that, due to the grasping rigidity, it holds that $\boldsymbol{\omega}_i=\boldsymbol{\omega}^{\scriptscriptstyle O}_j, \forall i\in\mathcal{T}$,
one obtains
\begin{equation}
\boldsymbol{v}_i=\boldsymbol{J}^{\scriptscriptstyle O}_{i,j}(\boldsymbol{q}_i) \boldsymbol{v}^{\scriptscriptstyle O}_j, \label{eq:J_o_i}
\end{equation}
where $\boldsymbol{J}^{\scriptscriptstyle O}_{i,j}:\mathbb{R}^{n_i}\to\mathbb{R}^{6\times6}$ is the object-to-agent Jacobian matrix, with
\begin{equation*}
\boldsymbol{J}^{\scriptscriptstyle O}_{i,j}(\boldsymbol{x})\coloneqq\left[\begin{array}{cc}
\boldsymbol{I}_{ 3} & -\boldsymbol{S}(\boldsymbol{R}_i(\boldsymbol{x}) \boldsymbol{p}^{\scriptscriptstyle E_i}_{\scriptscriptstyle E_{i}/O_j})\\
\boldsymbol{0}_{3\times 3} & \boldsymbol{I}_{3}
\end{array}\right], \forall \boldsymbol{x}\in\mathbb{R}^{n_i}, \label{eq:J_o_i_def}
\end{equation*}
which is always full-rank. 
The agent task-space dynamics \eqref{eq:manipulator dynamics} can be written in vector form as:
\begin{equation}
\boldsymbol{M}_{\scriptscriptstyle \mathcal{T}}(\boldsymbol{q}_{\scriptscriptstyle \mathcal{T}})\dot{\boldsymbol{v}}_{\scriptscriptstyle \mathcal{T}}+\boldsymbol{C}_{\scriptscriptstyle \mathcal{T}}(\boldsymbol{q}_{\scriptscriptstyle \mathcal{T}},\dot{\boldsymbol{q}}_{\scriptscriptstyle \mathcal{T}})\boldsymbol{v}_{\scriptscriptstyle \mathcal{T}} + \boldsymbol{g}_{\scriptscriptstyle \mathcal{T}}(\boldsymbol{q}_{\scriptscriptstyle \mathcal{T}})  = \boldsymbol{u}_{\scriptscriptstyle \mathcal{T}}-\boldsymbol{f}_{\scriptscriptstyle \mathcal{T}},\label{eq:manipulator dynamics_vector_form}
\end{equation}
where $\boldsymbol{q}_{\scriptscriptstyle \mathcal{T}} \coloneqq [ \boldsymbol{q}^\top_i]^\top_{\scriptscriptstyle i\in\mathcal{T}}$, $\dot{\boldsymbol{q}}_{\scriptscriptstyle \mathcal{T}}\coloneqq [ \dot{\boldsymbol{q}}^\top_i ]^\top_{ \scriptscriptstyle i\in\mathcal{T}}$, $\boldsymbol{v}_{\scriptscriptstyle \mathcal{T}} \coloneqq [ \boldsymbol{v}^\top_i]^\top_{\scriptscriptstyle i\in\mathcal{T}}$, $\boldsymbol{g}_{\scriptscriptstyle \mathcal{T}}(\boldsymbol{q}_{\scriptscriptstyle \mathcal{T}}) \coloneqq \Big[ [\boldsymbol{g}_i(\boldsymbol{q}_i)]^\top\Big]^\top_{\scriptscriptstyle i\in\mathcal{T}}$, $\boldsymbol{f}_{\scriptscriptstyle \mathcal{T}} \coloneqq [ \boldsymbol{f}^\top_i ]^\top_{ \scriptscriptstyle i\in\mathcal{T}}$ and  $\boldsymbol{M}_{\scriptscriptstyle \mathcal{T}}(\boldsymbol{q}_{\scriptscriptstyle \mathcal{T}})\coloneqq \text{diag}\{ [\boldsymbol{M}_i(\boldsymbol{q}_i)]_{\scriptscriptstyle i\in\mathcal{T}} \}$, $\boldsymbol{C}_{\scriptscriptstyle \mathcal{T}}(\boldsymbol{q}_{\scriptscriptstyle \mathcal{T}},\dot{\boldsymbol{q}}_{\scriptscriptstyle \mathcal{T}}) \coloneqq \text{diag}\{ [\boldsymbol{C}_i(\boldsymbol{q}_i,\dot{\boldsymbol{q}}_i)]_{i\in\mathcal{T}} \}$.

The kineto-statics duality along with the grasp rigidity suggest that the force $\boldsymbol{f}^{\scriptscriptstyle O}_j$ acting on the object's center of mass and the generalized forces $\boldsymbol{f}_i,i\in\mathcal{T}$, exerted by the agents at the grasping points, are related through:
\begin{equation}
\boldsymbol{f}^{\scriptscriptstyle O}_j=[\boldsymbol{G}_{\scriptscriptstyle \mathcal{T},j}(\boldsymbol{q}_{\scriptscriptstyle \mathcal{T}})]^\top\boldsymbol{f}_{\scriptscriptstyle \mathcal{T}},\label{eq:grasp matrix}
\end{equation}
where $\boldsymbol{G}_{\scriptscriptstyle \mathcal{T},j}:\mathbb{R}^{n_{\scriptscriptstyle \mathcal{T}}}\to\mathbb{R}^{6N\times 6}$,
with $\boldsymbol{G}_{\scriptscriptstyle \mathcal{T},j}(\boldsymbol{q}_{\scriptscriptstyle \mathcal{T}}) \coloneqq \Big[ [\boldsymbol{J}^{\scriptscriptstyle O}_{i,j}(\boldsymbol{q}_i)]^\top \Big]^\top_{\scriptscriptstyle i\in\mathcal{T}}$
is the grasp matrix, and $n_{\scriptscriptstyle \mathcal{T}} \coloneqq \sum_{i\in\mathcal{T}}n_i$. By combining \eqref{eq:grasp matrix} with \eqref{eq:object dynamics}, \eqref{eq:manipulator dynamics_vector_form}, \eqref{eq:J_o_i} and \eqref{eq:coupled_kinematics}  we obtain the coupled dynamics \cite{verginis2017ifacCoop_Manp}
\begin{align}
&\widetilde{\boldsymbol{M}}_{\scriptscriptstyle \mathcal{T},j}(\boldsymbol{x}_{\scriptscriptstyle \mathcal{T},j})\dot{\boldsymbol{v}}^{\scriptscriptstyle O}_j+\widetilde{\boldsymbol{C}}_{\scriptscriptstyle \mathcal{T},j}(\boldsymbol{x}_{\scriptscriptstyle \mathcal{T},j})\boldsymbol{v}^{\scriptscriptstyle O}_j+\widetilde{\boldsymbol{g}}_{\scriptscriptstyle \mathcal{T},j}(\boldsymbol{x}_{\scriptscriptstyle \mathcal{T},j})  = \notag \\
&\hspace{55mm} [\boldsymbol{G}_{\scriptscriptstyle \mathcal{T},j}(\boldsymbol{q}_{\scriptscriptstyle \mathcal{T}})]^\top\boldsymbol{u}_{\scriptscriptstyle \mathcal{T}},\label{eq:coupled dynamics}
\end{align}
where 
\small
\begin{align*}
&\widetilde{\boldsymbol{M}}_{\scriptscriptstyle \mathcal{T},j}(\boldsymbol{x}_{\scriptscriptstyle \mathcal{T},j})  \coloneqq  \boldsymbol{M}_{\scriptscriptstyle O}(\boldsymbol{x}^{\scriptscriptstyle O}_j)+[\boldsymbol{G}_{\scriptscriptstyle \mathcal{T},j}(\boldsymbol{q}_{\scriptscriptstyle \mathcal{T}})]^\top \boldsymbol{M}_{\scriptscriptstyle \mathcal{T}}(\boldsymbol{q}_{\scriptscriptstyle \mathcal{T}}) \boldsymbol{G}_{\scriptscriptstyle \mathcal{T},j}(\boldsymbol{q}_{\scriptscriptstyle \mathcal{T}}) \\
&\widetilde{\boldsymbol{C}}_{\scriptscriptstyle \mathcal{T},j}(\boldsymbol{x}_{\scriptscriptstyle \mathcal{T},j})  \coloneqq  \boldsymbol{C}_{\scriptscriptstyle O}(\boldsymbol{x}^{\scriptscriptstyle O}_j, \boldsymbol{v}^{\scriptscriptstyle O}_j)+ [\boldsymbol{G}_{\scriptscriptstyle \mathcal{T},j}(\boldsymbol{q}_{\scriptscriptstyle \mathcal{T}})]^\top \boldsymbol{M}_{\scriptscriptstyle \mathcal{T}}(\boldsymbol{q}_{\scriptscriptstyle \mathcal{T}})\dot{\boldsymbol{G}}_{\scriptscriptstyle \mathcal{T},j}(\boldsymbol{q}_{\scriptscriptstyle \mathcal{T}}) \notag \\ 
& \hspace{20mm} + [\boldsymbol{G}_{\scriptscriptstyle \mathcal{T},j}(\boldsymbol{q}_{\scriptscriptstyle \mathcal{T}})]^\top \boldsymbol{C}_{\scriptscriptstyle \mathcal{T}}(\boldsymbol{q}_{\scriptscriptstyle \mathcal{T}},\dot{\boldsymbol{q}}_{\scriptscriptstyle \mathcal{T}})\boldsymbol{G}_{\scriptscriptstyle \mathcal{T},j}(\boldsymbol{q}_{\scriptscriptstyle \mathcal{T}}) \\
&\widetilde{\boldsymbol{g}}_{\scriptscriptstyle \mathcal{T},j}(\boldsymbol{x}_{\scriptscriptstyle \mathcal{T},j})  \coloneqq  \boldsymbol{g}_{\scriptscriptstyle O}(\boldsymbol{x}^{\scriptscriptstyle O}_j)+[\boldsymbol{G}_{\scriptscriptstyle \mathcal{T},j}(\boldsymbol{q}_{\scriptscriptstyle \mathcal{T}})]^\top \boldsymbol{g}_{\scriptscriptstyle \mathcal{T}}(\boldsymbol{q}_{\scriptscriptstyle \mathcal{T}}).
\end{align*}
\normalsize
and $\boldsymbol{x}_{\scriptscriptstyle \mathcal{T},j}$ is the overall state $\boldsymbol{x}_{\scriptscriptstyle \mathcal{T},j} \coloneqq [\boldsymbol{q}_{\scriptscriptstyle \mathcal{T}}^\top,\dot{\boldsymbol{q}}_{\scriptscriptstyle \mathcal{T}}^\top, (\boldsymbol{x}^{\scriptscriptstyle O}_j)^\top, ({\boldsymbol{v}}^{\scriptscriptstyle O}_j)^\top]^\top\in\mathbb{R}^{2n_{\scriptscriptstyle \mathcal{T}}+6}\times\mathbb{M}$. Note that the aforementioned coupled terms are defined only when $\boldsymbol{q}_i\in\mathbb{S}_i\subset \mathbb{R}^{n_i}, \forall i\in\mathcal{T}$.
Moreover, the following Lemma is necessary for the following analysis. 
\begin{lemma} \label{lem:coupled dynamics skew symmetry}
\cite{verginis2017ifacCoop_Manp} The matrices $\boldsymbol{B}_i(\boldsymbol{q}_i)$ and $\widetilde{\boldsymbol{M}}_{\scriptscriptstyle \mathcal{T},j}(\boldsymbol{x}_{\scriptscriptstyle \mathcal{T},j})$ are symmetric and positive definite and the matrices $\dot{\boldsymbol{B}}_i(\boldsymbol{q}_i) - 2\boldsymbol{N}_i(\boldsymbol{q}_i,\dot{\boldsymbol{q}}_i)$ and $\dot{\widetilde{\boldsymbol{M}}}_{\scriptscriptstyle \mathcal{T},j}(\boldsymbol{x}_{\scriptscriptstyle \mathcal{T},j}) - 2\widetilde{\boldsymbol{C}}_{\scriptscriptstyle \mathcal{T},j}(\boldsymbol{x}_{\scriptscriptstyle \mathcal{T},j})$ are skew symmetric, $\forall i\in\mathcal{N}, j\in\mathcal{M}, \mathcal{T} \subseteq\mathcal{N}$.
\end{lemma}

Regarding the volume of the coupled agents-object system, we denote by $\mathcal{AO}_{\mathcal{T},j}:\mathbb{R}^3 \rightrightarrows \mathbb{R}^3$ the sphere centered at $\boldsymbol{p}^{\scriptscriptstyle O}_j$ with constant radius $r_{\scriptscriptstyle \mathcal{T},j} \in\mathbb{R}_{>0}$, i.e., $\mathcal{AO}_{\mathcal{T},j}(\boldsymbol{p}^{\scriptscriptstyle O}_j) \coloneqq \mathcal{B}(\boldsymbol{p}^{\scriptscriptstyle O}_j,r_{\scriptscriptstyle \mathcal{T},j})$, which is large enough to cover the volume of the coupled system in all configurations $\boldsymbol{q}_{\scriptscriptstyle \mathcal{T}}$\footnote{$r_{\scriptscriptstyle \mathcal{T},j}$ can be chosen as the largest distance of the object's center of mass to a point in the agents' volume over all possible $\boldsymbol{q}_{\scriptscriptstyle \mathcal{T}}$ (see \cite{verginis2017distributed},  Section IIIB.)}. This conservative formulation emanates from the sphere-world restriction of the multi-agent navigation function framework \cite{Loizou2006,koditschek1990robot}. In order to take into account other spaces, ideas from \cite{Loizou-RSS-14} could be employed or extensions of the respective works of \cite{rimon1992exact}, \cite{loizou2017navigation} to the multi-agent case could be developed.

Moreover, in order to take into account the introduced agents' power capabilities $\zeta_i$, $i\in\mathcal{N}$, we consider a function $\Lambda\in\{\top,\bot\}$ that outputs whether the agents that grasp an object are able to transport the object, based on their power capabilities. For instance, $\Lambda(m^{\scriptscriptstyle O}_j, \zeta_{\mathcal{T}}) = \top$, where $m^{\scriptscriptstyle O}_j \in\mathbb{R}_{>0}$ is the mass of object $j$ and $\zeta_{\mathcal{T}} \coloneqq [\zeta_i]^\top_{i\in\mathcal{T}}$, implies that the agents $\mathcal{T}$ have sufficient power capabilities to cooperatively transport object $j$. 

\subsection{Problem Formulation} \label{subsec:PF}

In this subsection, the problem formulation is provided. We first introduce some preliminary required notation. 
We define the boolean functions $\mathcal{AG}_{i,j}:\mathbb{R}^{n_i}\times\mathbb{M}\to\{\top,\bot\}, i\in\mathcal{N},j\in\mathcal{M}$, to denote whether agent $i\in\mathcal{N}$ rigidly grasps an object $j\in\mathcal{M}$ at a given configuration $\boldsymbol{q}_i,\boldsymbol{x}^{\scriptscriptstyle O}_j$; We also define $\mathcal{AG}_{i,0}: \mathbb{R}^{n_i}\times\mathbb{M}^M\to\{\top,\bot\}$, to denote that agent $i$ does not grasp any objects, i.e., $\mathcal{AG}_{i,j}(\boldsymbol{q}_i,\boldsymbol{x}^{\scriptscriptstyle O}_j) =  \bot, \forall j\in\mathcal{M} \Leftrightarrow \mathcal{AG}_{i,0}(\boldsymbol{q}_i, \boldsymbol{x}^{\scriptscriptstyle O}) = \top$, $\forall i\in\mathcal{N}$, where $\boldsymbol{x}^{\scriptscriptstyle O} \coloneqq [ (\boldsymbol{x}^{\scriptscriptstyle O}_j)^\top]^\top_{j\in\mathcal{M}}\in\mathbb{M}^M$.
Note also that $\mathcal{AG}_{i,\ell}(\boldsymbol{q}_i,\boldsymbol{x}^{\scriptscriptstyle O}_\ell) =\top, \ell\in\mathcal{M} \Leftrightarrow \mathcal{AG}_{i,j}(\boldsymbol{q}_i,\boldsymbol{x}^{\scriptscriptstyle O}_j)=\bot, \forall j\in \mathcal{M}\backslash\{\ell\}$, i.e., agent $i$ can grasp at most one object at a time.

In addition, we use the boolean functions $\mathcal{C}_{i,l}: \mathbb{R}^{n_i+n_l}\to\{\bot,\top\}$, $\mathcal{C}_{i,\scriptscriptstyle O_j}: \mathbb{R}^{n_i}\times\mathbb{M}\to\{\bot,\top\}$, $\mathcal{C}_{\scriptscriptstyle O_j,\scriptscriptstyle O_\ell}: \mathbb{M}^2\to\{\bot,\top\}$, to denote collision between agents $i,l\in\mathcal{N}, i\neq l$, agent $i\in\mathcal{N}$ and object $j\in\mathcal{M}$ and objects $j,\ell\in\mathcal{M}, j\neq \ell$, respectively. 

We also assume the  existence of a procedure $\boldsymbol{\mathcal{P}}_s$ that outputs whether or not a set of non-intersecting spheres fits in a larger sphere as well as possible positions of the spheres in the case they fit. More specifically, given a region of interest $\pi_k$ and a number $\widetilde{N}\in\mathbb{N}$ of sphere radii (of agents and/or objects) the procedure can be seen as a function $\boldsymbol{\mathcal{P}}_s \coloneqq [\mathcal{P}_{s,0}, \boldsymbol{\mathcal{P}}^\top_{s,1}]^\top$, where $\mathcal{P}_{s,0}:\mathbb{R}^{\widetilde{N}+1}_{\geq 0}\to\{\top,\bot\}$ outputs whether the spheres fit in the region $\pi_k$ whereas  $\boldsymbol{\mathcal{P}}_{s,1}$ provides possible configurations of the agents and the objects 
or $\boldsymbol{0}$ in case the spheres do not fit.
For instance, $P_{s,0}(r_{\pi_2},r_1,r_3,r^{\scriptscriptstyle O}_1,r^{\scriptscriptstyle O}_5)$ determines whether the agents $1,3$ and the objects $1,5$ fit in region $\pi_2$, without colliding with each other; $(\boldsymbol{q}_1,\boldsymbol{q}_3,\boldsymbol{x}^{\scriptscriptstyle O}_1, \boldsymbol{x}^{\scriptscriptstyle O}_5)=\boldsymbol{P}_{s,1}(r_{\pi_2},r_1,r_3,r^{\scriptscriptstyle O}_1,r^{\scriptscriptstyle O}_5)$ provides a set of configurations such that $\mathcal{A}_1(\boldsymbol{q}_1), \mathcal{A}_3(\boldsymbol{q}_3), \mathcal{O}_1(\boldsymbol{x}^{\scriptscriptstyle O}_1), \mathcal{O}_{5}(\boldsymbol{x}^{\scriptscriptstyle O}_5) \subset \pi_2$ and $\mathcal{C}_{1,3}(\boldsymbol{q}_1,\boldsymbol{q}_3) = \mathcal{C}_{\scriptscriptstyle O_1,O_5}(\boldsymbol{x}^{\scriptscriptstyle O}_1,\boldsymbol{x}^{\scriptscriptstyle O}_5) = \mathcal{C}_{i,\scriptscriptstyle O_j}(\boldsymbol{q}_i,\boldsymbol{x}^{\scriptscriptstyle O}_j) = \bot, \forall (i,j)\in\{1,3\}\times\{1,5\}$. The problem of finding an algorithm $\mathcal{P}_s$ is a special case of the sphere packing problem \cite{chen2008sphere}. Note, however, that we are not interested in finding the maximum number of spheres that can be packed in a larger sphere but, rather, in the simpler problem of determining whether a set of spheres can be packed in a larger sphere. 

The following definitions address the transitions of the agents and the objects between the regions of interest. 

\begin{definition}(\textbf{Transition})  \label{def:agent transition}
Consider that $\mathcal{A}_i(\boldsymbol{q}_i(t_0)) \subset \pi_k$, for some $i\in\mathcal{N}, k\in\mathcal{K}, t_0\in\mathbb{R}_{\geq 0}$, and $\mathcal{C}_{i,l}(\boldsymbol{q}_i(t_0),\boldsymbol{q}_l(t_0)) = \mathcal{C}_{i,\scriptscriptstyle O_j}(\boldsymbol{q}_j(t_0),\boldsymbol{x}^{\scriptscriptstyle O}_j(t_0)) = \bot, \forall l\in\mathcal{N}\backslash\{i\},j\in\mathcal{M}$.  Then, there exists a transition for agent $i$ from region $\pi_k$ to $\pi_{k'},k'\in\mathcal{K}$, denoted as $\pi_k\rightarrow_i\pi_{k'}$, if there exists a finite $t_f\geq t_0$ and a bounded feedback control trajectory $\boldsymbol{u}_i$ such that $\mathcal{A}_i(\boldsymbol{q}_i(t_f))\subset\pi_{k'}$, $\mathcal{C}_{i,l}(\boldsymbol{q}_i(t),\boldsymbol{q}_l(t)) = \mathcal{C}_{i,\scriptscriptstyle O_j}(\boldsymbol{q}_i(t),\boldsymbol{x}^{\scriptscriptstyle O}_j(t)) = \bot$, and $\mathcal{A}_i(\boldsymbol{q}_i(t))\cap\pi_m = \emptyset$, $\forall t\in[t_0,t_f], l\in\mathcal{N}\backslash\{i\},j\in\mathcal{M}, m\in\mathcal{K}\backslash\{k,k'\}$.
\end{definition}

\begin{definition}(\textbf{Grasping}) \label{def:grasping }
Consider that $\mathcal{A}_i(\boldsymbol{q}_i(t_0))\subset\pi_k$, $\mathcal{O}_j(\boldsymbol{x}^{\scriptscriptstyle O}_j(t_0))\subset\pi_k$, $k\in\mathcal{K}$ for some $i\in\mathcal{N}$, $j\in\mathcal{M}$, $t_0\in\mathbb{R}_{\geq 0}$, with $\mathcal{AG}_{i,0}(\boldsymbol{q}_i(t_0),\boldsymbol{x}^{\scriptscriptstyle O}(t_0)) = \top$, and 
\begin{enumerate}
\item $\mathcal{C}_{i,l}(\boldsymbol{q}_i(t_0),\boldsymbol{q}_l(t_0)) = \mathcal{C}_{i,\scriptscriptstyle O_{j'}}(\boldsymbol{q}_i(t_0),\boldsymbol{x}^{\scriptscriptstyle O}_{j'}(t_0)) = \bot, \forall l\in\mathcal{N}\backslash\{i\}, j' \in\mathcal{M}$,
\item $\mathcal{C}_{i',\scriptscriptstyle O_j}(\boldsymbol{q}_{i'}(t_0),\boldsymbol{x}^{\scriptscriptstyle O}_j(t_0)) =  \mathcal{C}_{\scriptscriptstyle O_j,\scriptscriptstyle O_\ell}(\boldsymbol{x}^{\scriptscriptstyle O}_j(t_0),\boldsymbol{x}^{\scriptscriptstyle O}_\ell(t_0)) = \bot, \forall i'\in\mathcal{N}, \ell \in\mathcal{M}\backslash\{j\}$. 
\end{enumerate}  
Then, agent $i$ \textit{grasps} object $j$, denoted as $i\xrightarrow{g}j$, if there exists a finite $t_f\geq t_0$ and a bounded control trajectory $\boldsymbol{u}_i$ such that $\mathcal{AG}_{i,j}(\boldsymbol{q}_i(t_f),\boldsymbol{x}^{\scriptscriptstyle O}_j(t_f)) = \top$, $\mathcal{A}_i(\boldsymbol{q}_i(t))\subset\pi_k$,   $\mathcal{O}_j(\boldsymbol{x}^{\scriptscriptstyle O}_j(t))\subset\pi_k$, $k\in\mathcal{K}$ with 
\begin{enumerate}
\item $\mathcal{C}_{i,l}(\boldsymbol{q}_i(t),\boldsymbol{q}_l(t)) = \mathcal{C}_{i,\scriptscriptstyle O_{\ell}}(\boldsymbol{q}_i(t),\boldsymbol{x}^{\scriptscriptstyle O}_{\ell}(t)) = \bot$,
\item $\mathcal{C}_{l,\scriptscriptstyle O_j}(\boldsymbol{q}_{l}(t),\boldsymbol{x}^{\scriptscriptstyle O}_j(t)) =  \mathcal{C}_{\scriptscriptstyle O_j,\scriptscriptstyle O_\ell}(\boldsymbol{x}^{\scriptscriptstyle O}_j(t),\boldsymbol{x}^{\scriptscriptstyle O}_\ell(t)) = \bot$, 
\end{enumerate}  
$\forall t\in[t_0,t_f], l\in\mathcal{N}\backslash\{i\}, \ell \in\mathcal{M}\backslash\{j\}$.
\end{definition}

\begin{definition}(\textbf{Releasing}) \label{def:releasing }
Consider that $\mathcal{A}_i(\boldsymbol{q}_i(t_0))\subset\pi_k$, $\mathcal{O}_j(\boldsymbol{x}^{\scriptscriptstyle O}_j(t_0))\subset\pi_k$, $k\in\mathcal{K}$ for some $i\in\mathcal{N}$, $j\in\mathcal{M}$, $t_0\in\mathbb{R}_{\geq 0}$, with $\mathcal{AG}_{i,j}(\boldsymbol{q}_i(t_0),\boldsymbol{x}^{\scriptscriptstyle O}_j(t_0)) = \top$, and 
\begin{enumerate}
\item $\mathcal{C}_{i,l}(\boldsymbol{q}_i(t_0),\boldsymbol{q}_l(t_0)) = \mathcal{C}_{i,\scriptscriptstyle O_{\ell}}(\boldsymbol{q}_i(t_0),\boldsymbol{x}^{\scriptscriptstyle O}_{\ell}(t_0)) = \bot$,
\item $\mathcal{C}_{l,\scriptscriptstyle O_j}(\boldsymbol{q}_{l}(t_0),\boldsymbol{x}^{\scriptscriptstyle O}_j(t_0)) =  \mathcal{C}_{\scriptscriptstyle O_j,\scriptscriptstyle O_\ell}(\boldsymbol{x}^{\scriptscriptstyle O}_j(t_0),\boldsymbol{x}^{\scriptscriptstyle O}_\ell(t_0)) = \bot$, 
\end{enumerate}  
$\forall l\in\mathcal{N}\backslash\{i\}, \ell \in\mathcal{M}\backslash\{j\}$.
Then, agent $i$ \textit{releases} object $j$, denoted as $i\xrightarrow{r}j$, if there exists a finite $t_f\geq t_0$ and a bounded control trajectory $\boldsymbol{u}_i$ such that $\mathcal{AG}_{i,0}(\boldsymbol{q}_i(t_f),\boldsymbol{x}^{\scriptscriptstyle O}(t_f)) = \top$, $\mathcal{A}_i(\boldsymbol{q}_i(t))\subset\pi_k$,  $\mathcal{O}_j(\boldsymbol{x}^{\scriptscriptstyle O}_j(t))\subset\pi_k$, $k\in\mathcal{K}$ with 
\begin{enumerate}
\item $\mathcal{C}_{i,l}(\boldsymbol{q}_i(t),\boldsymbol{q}_l(t)) = \mathcal{C}_{i,\scriptscriptstyle O_{\ell}}(\boldsymbol{q}_i(t),\boldsymbol{x}^{\scriptscriptstyle O}_{\ell}(t)) = \bot$,
\item $\mathcal{C}_{l,\scriptscriptstyle O_j}(\boldsymbol{q}_{l}(t),\boldsymbol{x}^{\scriptscriptstyle O}_j(t)) =  \mathcal{C}_{\scriptscriptstyle O_j,\scriptscriptstyle O_\ell}(\boldsymbol{x}^{\scriptscriptstyle O}_j(t),\boldsymbol{x}^{\scriptscriptstyle O}_\ell(t)) = \bot$, 
\end{enumerate}  
$\forall t\in[t_0,t_f], l\in\mathcal{N}\backslash\{i\}, \ell \in\mathcal{M}\backslash\{j\}$.
\end{definition}

\begin{definition}(\textbf{Transportation}) \label{def:agent-object transition}
Consider a nonempty subset of agents $\mathcal{T}\subseteq\mathcal{N}$ with $\mathcal{A}_i(\boldsymbol{q_i}(t_0))\subset\pi_k$, $\forall i\in\mathcal{T}$, and $\mathcal{O}_j(\boldsymbol{x}^{\scriptscriptstyle O}_j(t_0))\subset\pi_k$, for some $j\in\mathcal{M}$, $k\in\mathcal{K}$, $t_0\geq 0$, with $\mathcal{AG}_{i,j}(\boldsymbol{q}_i(t_0), \boldsymbol{x}^{\scriptscriptstyle O}_j(t_0)) = \top$, $\forall i\in\mathcal{T}$ and
\begin{enumerate}
\item $\mathcal{C}_{i,l}(\boldsymbol{q}_i(t_0),\boldsymbol{q}_l(t_0)) = \mathcal{C}_{i,\scriptscriptstyle O_{\ell}}(\boldsymbol{q}_i(t_0),\boldsymbol{x}^{\scriptscriptstyle O}_{\ell}(t_0)) = \bot$, 
\item $\mathcal{C}_{z,\scriptscriptstyle O_j}(\boldsymbol{q}_{l}(t_0),\boldsymbol{x}^{\scriptscriptstyle O}_j(t_0)) =  \mathcal{C}_{\scriptscriptstyle O_j,\scriptscriptstyle O_\ell}(\boldsymbol{x}^{\scriptscriptstyle O}_j(t_0),\boldsymbol{x}^{\scriptscriptstyle O}_\ell(t_0)) = \bot$, 
\end{enumerate}  
$\forall i,l\in\mathcal{N}$, with $i\neq l$, $\ell \in\mathcal{M}\backslash\{j\}, z\in\mathcal{N}\backslash\mathcal{T}$.
Then, the team of agents $\mathcal{T}$ transports the object $j$ from region $\pi_k$ to region $\pi_{k'}, k'\in\mathcal{K}$, denoted as $\pi_k \xrightarrow{T}_{\mathcal{T},j} \pi_{k'}$, if there exists a finite $t_f \geq t_0$ and bounded control laws $\boldsymbol{u}_i, i\in\mathcal{T}$, such that $\mathcal{A}_i(\boldsymbol{q}_i(t_f))\subset\pi_{k'},\forall i\in\mathcal{T}, \mathcal{O}_j(\boldsymbol{x}^{\scriptscriptstyle O}_j(t_f))\subset\pi_{k'}$, $\mathcal{AG}_{i,j}(\boldsymbol{q}_i(t),\boldsymbol{x}^{\scriptscriptstyle O}_j(t)) = \top$, and

\begin{enumerate}
\item $\mathcal{C}_{i,l}(\boldsymbol{q}_i(t),\boldsymbol{q}_l(t)) = \mathcal{C}_{i,\scriptscriptstyle O_{\ell}}(\boldsymbol{q}_i(t),\boldsymbol{x}^{\scriptscriptstyle O}_{\ell}(t)) = \bot$,
\item $\mathcal{C}_{z,\scriptscriptstyle O_j}(\boldsymbol{q}_{l}(t),\boldsymbol{x}^{\scriptscriptstyle O}_j(t)) =  \mathcal{C}_{\scriptscriptstyle O_j,\scriptscriptstyle O_\ell}(\boldsymbol{x}^{\scriptscriptstyle O}_j(t),\boldsymbol{x}^{\scriptscriptstyle O}_\ell(t)) = \bot$, 
\end{enumerate}  
and $\mathcal{AO}_{\mathcal{T},j}(\boldsymbol{p}^{\scriptscriptstyle O}_j(t))\cap \pi_m = \emptyset$, $\forall t\in[t_0,t_f], i,l\in\mathcal{N}$, with $i\neq l$, $\ell \in\mathcal{M}\backslash\{j\}, z\in\mathcal{N}\backslash\mathcal{T}, m\in\mathcal{K}\backslash\{k,k'\}$.
\end{definition}
Loosely speaking, the aforementioned definitions correspond to specific actions of the agents, namely \textit{transition}, \textit{grasp}, \textit{release}, and \textit{transport}. We do not define these actions explicitly though, since we will employ directly designed continuous control inputs $u_i$, as will be seen later. Moreover, in the \textit{grasping/releasing} definitions, we have not incorporated explicitly collisions between the agent and the object to be grasped/released other than the grasping point. Such collisions will be assumed to be avoided in the next section.    


Our goal is to control the multi-agent system such that the agents and the objects obey a given specification over their atomic propositions $\Psi_i, \Psi^{\scriptscriptstyle O}_j, \forall i\in\mathcal{N},j\in\mathcal{M}$. 
Given the trajectories $\boldsymbol{q}_i(t), \boldsymbol{x}^{\scriptscriptstyle O}_j(t), t\in\mathbb{R}_{\geq 0}$, of agent $i$ and object $j$, respectively, their corresponding \textit{behaviors} are given by the infinite sequences 
\begin{align*}
& b_i \coloneqq (\boldsymbol{q}_i(t),\sigma_i) \coloneqq (\boldsymbol{q}_i(t_{i,1}),\sigma_{i,1})(\boldsymbol{q}_i(t_{i,2}),\sigma_{i,2})\dots, \\ 
& b^{\scriptscriptstyle O}_j \coloneqq (\boldsymbol{x}^{\scriptscriptstyle O}_j(t),\sigma^{\scriptscriptstyle O}_j) \coloneqq (\boldsymbol{x}^{\scriptscriptstyle O}_j(t^{\scriptscriptstyle O}_{j,1}),\sigma^{\scriptscriptstyle O}_{j,1}) (\boldsymbol{x}^{\scriptscriptstyle O}_j(t^{\scriptscriptstyle O}_{j,2}),\sigma^{\scriptscriptstyle O}_{j,2})\dots,
\end{align*}
 with $t_{i,\ell+1} > t_{i,\ell} \geq 0, t^{\scriptscriptstyle O}_{j,\ell+1} > t^{\scriptscriptstyle O}_{j,\ell} \geq 0, \forall \ell\in\mathbb{N}$, representing specific time stamps. The sequences $\sigma_i, \sigma^{\scriptscriptstyle O}_j$ are the services provided to the agent and the object, respectively, over their trajectories, i.e., $\sigma_{i,\ell}\in 2^{\Psi_i}, \sigma^{\scriptscriptstyle O}_{j,l}\in 2^{\Psi^{\scriptscriptstyle O}_j}$ with $\mathcal{A}_i(\boldsymbol{q}_i(t_{i,\ell})) \subset \pi_{k_{i,\ell}}, \sigma_{i,\ell} \in \mathcal{L}_i(\pi_{k_{i,\ell}})$ and $\mathcal{O}_j(\boldsymbol{x}^{\scriptscriptstyle O}_j(t^{\scriptscriptstyle O}_{j,l})) \subset \pi_{k^{\scriptscriptstyle O}_{j,l}}, \sigma^{\scriptscriptstyle O}_{j,l} \in \mathcal{L}^{\scriptscriptstyle O}_j(\pi_{k^{\scriptscriptstyle O}_{j,l}}), k_{i,\ell}, k^{\scriptscriptstyle O}_{j,l} \in\mathcal{K}, \forall \ell,l\in\mathbb{N}, i\in\mathcal{N}, j\in\mathcal{M}$, where $\mathcal{L}_i$ and $\mathcal{L}^{\scriptscriptstyle O}_j$ are the previously defined labeling functions. The following Lemma then follows: 
\begin{lemma}
The behaviors $b_i, b^{\scriptscriptstyle O}_j$ satisfy formulas $\phi_i, \phi_{\scriptscriptstyle O_j}$ if $\sigma_i \models \phi_i$ and $\sigma^{\scriptscriptstyle O}_j \models \phi^{\scriptscriptstyle O}_j$, respectively.
\end{lemma}

The control objectives are given as LTL formulas $\phi_i, \phi^{\scriptscriptstyle O}_j$ over $\Psi_i, \Psi^{\scriptscriptstyle O}_j$, respectively, $\forall i\in\mathcal{N},j\in\mathcal{M}$. The LTL formulas $\phi_i, \phi^{\scriptscriptstyle O}_j$ are satisfied if there exist behaviors $b_i,b^{\scriptscriptstyle O}_j$ of agent $i$ and object $j$ that satisfy  $\phi_i, \phi^{\scriptscriptstyle O}_j$. We are now ready to give a formal problem statement:
\begin{prob} \label{problem}
Consider $N$ robotic agents and $M$ objects in $\mathcal{W}$ subject to the dynamics \eqref{eq:manipulator dynamics} and \eqref{eq:object dynamics}, respectively, and 
\begin{enumerate}
\item  $\dot{\boldsymbol{q}}_i(0) = \boldsymbol{0}, \boldsymbol{v}^{\scriptscriptstyle O}_{j} = \boldsymbol{0}$, $\mathcal{A}_i(\boldsymbol{q}_i(0))\subset \pi_{\text{init}(i)}, \mathcal{O}_j(\boldsymbol{x}^{\scriptscriptstyle O}_j(0)) \subset\pi_{\text{init}_{\scriptscriptstyle O}(j)}$, $\forall i\in\mathcal{N},j\in\mathcal{M}$, 
\item $\mathcal{C}_{i,l}(\boldsymbol{q}_i(0),\boldsymbol{q}_l(0)) = \mathcal{C}_{\scriptscriptstyle O_j,\scriptscriptstyle O_\ell}(\boldsymbol{x}^{\scriptscriptstyle O}_j(0),\boldsymbol{x}^{\scriptscriptstyle O}_\ell(0)) = \mathcal{C}_{i,\scriptscriptstyle O_j}(\boldsymbol{q}_{i}(0),\boldsymbol{x}^{\scriptscriptstyle O}_\ell(0)) =  \bot$, $\forall i,l\in\mathcal{N}, i\neq l, j,\ell\in\mathcal{M},j\neq \ell$.  
\end{enumerate}
Given the disjoint sets $\Psi_i,\Psi^{\scriptscriptstyle O}_j$, $N$ LTL formulas $\phi_i$ over $\Psi_i$ and $M$ LTL formulas $\phi^{\scriptscriptstyle O}_j$ over $\Psi^{\scriptscriptstyle O}_j$, develop a control strategy that achieves behaviors $b_i, b^{\scriptscriptstyle O}_j$ which yield the satisfaction of 
$\phi_i, \phi^{\scriptscriptstyle O}_j, \forall i\in\mathcal{N},j\in\mathcal{M}$.
\end{prob}
Note that it is implicit in the problem statement the fact that the agents/objects starting in the same region can actually fit without colliding with each other. Technically, it holds that $\mathcal{P}_{s,0}(r_{\pi_k},[r_i]_{i\in \{i\in\mathcal{N}: \text{init}(i)= k\} }, [r^{\scriptscriptstyle O}_j]_{j\in \{ j\in\mathcal{M}: \text{init}_{\scriptscriptstyle O}(j) = k\} }) = \top$, $\forall k\in\mathcal{K}$.

\section{Main Results} \label{sec:main results}

\subsection{Continuous Control Design}
The first ingredient of our solution is the development of feedback control laws that establish agent transitions and object transportations as defined in Def. \ref{def:agent transition} and \ref{def:agent-object transition}, respectively. We do not focus on the grasping/releasing actions of Def. \ref{def:grasping }, \ref{def:releasing } and we refer to some existing methodologies that can derive the corresponding control laws (e.g., \cite{Cutkosky2012},\cite{Reis2015}). 

Assume that the conditions of Problem \ref{problem} hold for some $t_0 \in\mathbb{R}_{\geq 0}$, i.e., all agents and objects are located in regions of interest with zero velocity.
We design a control law such that a subset of agents performs a transition between two regions of interest and another subset of agents performs cooperative object transportation, according to Def. \ref{def:agent transition} and \ref{def:agent-object transition}, respectively. 
Let $\mathcal{Z}, \mathcal{T}, \mathcal{G}, \mathcal{R}\subseteq \mathcal{N}$ denote disjoint sets of agents corresponding to transition, transportation, grasping and releasing actions, respectively, with $\lvert \mathcal{Z} \rvert + |\mathcal{T}| + \lvert \mathcal{G} \rvert + \lvert \mathcal{R} \rvert  \leq \lvert \mathcal{N} \rvert$ and $\mathcal{A}_z(\boldsymbol{q}_z(t_0))\subset \pi_{k_z}$, $\mathcal{A}_{\tau}(\boldsymbol{q}_\tau(t_0))\subset \pi_{k_\tau}$, $\mathcal{A}_{g}(\boldsymbol{q}_g(t_0))\subset \pi_{k_g}$, $\mathcal{A}_\rho(\boldsymbol{q}_\rho(t_0))\subset \pi_{k_\rho}$, where $k_z,k_\tau,k_g,k_\rho\in\mathcal{K}$, $\forall z\in\mathcal{Z}, \tau\in\mathcal{T},g\in\mathcal{G},\rho\in\mathcal{R}$. Note that there might be idle agents in some regions, not performing any actions, i.e., the set $\mathcal{N}\backslash(\mathcal{Z}\cup\mathcal{V}\cup\mathcal{G}\cup\mathcal{Q})$ might not be empty.

More specifically, regarding the transportation actions, we consider that the set $\mathcal{T}$ consists of $\bar{T}$ disjoint teams of agents, with each team consisting of agents that are in the same region of interest and aim to collaboratively transport an object, i.e. $\mathcal{T} = \mathcal{T}_1\cup\mathcal{T}_2\cup\dots\mathcal{T}_{\bar{T}}$, and $\mathcal{A}_{\tau}(\boldsymbol{q}_{\tau}(t_0))\subset \pi_{k_{\mathcal{T}_m}}, \forall \tau\in\mathcal{T}_m, m\in\{1,\dots,\bar{T}\}$, where $k_{\mathcal{T}_m}\in\mathcal{K}, \forall m\in\{1,\dots,\bar{T}\}$.
Let also $\mathcal{S}\coloneqq\{s_{\mathcal{T}_1},s_{\mathcal{T}_2},\dots, s_{\mathcal{T}_{\bar{T}}}\}, \mathcal{X}\coloneqq\{[x_g]_{g\in\mathcal{G}}\}, \mathcal{Y}\coloneqq\{[y_\rho]_{\rho\in\mathcal{R}}\}\subseteq\mathcal{M}$ be disjoint sets of objects to be transported, grasped, and released, respectively. More specifically, each team $\mathcal{T}_m$ in the set $\mathcal{T}$ will transport cooperatively object $s_{\mathcal{T}_m}$, $m\in\{1,\dots,\bar{T}\}$, each agent $g\in\mathcal{G}$ will grasp object $x_g\in\mathcal{X}$ and each agent $\rho\in\mathcal{R}$ will release object $y_\rho\in\mathcal{Y}$. 
Then, suppose that the following conditions also hold at $t_0$:
\begin{itemize}
\item $\mathcal{O}_{s_{\mathcal{T}_m}}(\boldsymbol{x}^{\scriptscriptstyle O}_{s_{\mathcal{T}_m}}(t_0))\subset \pi_{k_{\mathcal{T}_m}}, \forall m\in\{1,\dots,\bar{T}\}$, $\mathcal{O}_{x_g}(\boldsymbol{x}^{\scriptscriptstyle O}_{x_g}(t_0))\subset \pi_{k_g}, \forall g\in\mathcal{G}$, \\ $\mathcal{O}_{y_\rho}(\boldsymbol{x}^{\scriptscriptstyle O}_{y_\rho}(t_0))\subset \pi_{k_\rho}, \forall \rho\in\mathcal{R}$,
\item $\mathcal{AG}_{\rho,y_\rho}(\boldsymbol{q}_\rho(t_0),\boldsymbol{x}^{\scriptscriptstyle O}_{y_\rho}(t_0)) = \top, \forall \rho\in\mathcal{R}$, $\mathcal{AG}_{z,0}(\boldsymbol{q}_z(t_0),\boldsymbol{x}^{\scriptscriptstyle O}(t_0)) = \top,\forall z\in\mathcal{Z}$, \\ $\mathcal{AG}_{g,0}(\boldsymbol{q}_g(t_0),\boldsymbol{x}^{\scriptscriptstyle O}(t_0))=\top, \forall g\in\mathcal{G}$, $\mathcal{AG}_{\tau,  s_{\mathcal{T}_m}}(\boldsymbol{q}_\tau(t_0),\boldsymbol{x}^{\scriptscriptstyle O}_{s_{\mathcal{T}_m}}(t_0))  = \top$, $\forall \tau\in\mathcal{T}_m, m\in\{1,\dots,\bar{T}\}$,
\end{itemize}
which mean, intuitively, that the objects $s_{\mathcal{T}_m}$, $x_g, y_\rho$ to be transported, grasped, released, are in the regions $\pi_{k_{\mathcal{T}_m}}$, $\pi_{k_g}$, $\pi_{k_\rho}$, respectively, and there is also grasping compliance with the corresponding agents.  
By also assuming that the agents do not collide with each other or with the objects (except for the transportation/releasing task agents), we guarantee that the conditions of Def. \ref{def:agent transition}-\ref{def:agent-object transition} hold. 

In the following, we design $\boldsymbol{u}_{z}$ and $\boldsymbol{u}_{\tau}$ such that $\pi_{k_z}\rightarrow_z\pi_{k'_z}$ and $\pi_{k_{\mathcal{T}_m}}\xrightarrow{T}_{\mathcal{T}_m,s_{\mathcal{T}_m}} \pi_{k'_{\mathcal{T}_m}}$, with $k'_z, k'_{\tau_m} \in\mathcal{K}, \forall z\in\mathcal{Z}, m\in\{1,\dots,\bar{T}\}$, assuming that (i) there exist appropriate $\boldsymbol{u}_g$ and $\boldsymbol{u}_\rho$ that guarantee $g\xrightarrow{g}x_g$ and $\rho\xrightarrow{r}y_\rho$ in $\pi_{k_g}, \pi_{k_\rho}$, respectively, $\forall g\in\mathcal{G},\rho\in\mathcal{R}$ and (ii) that the agents and objects fit in their respective goal regions, i.e., 
\begin{align}
&\mathcal{P}_{s,0}\Big(r_{\pi_k}, [r_z]_{z\in\mathcal{Q}_{\mathcal{Z},k}}, [r_g]_{g\in\mathcal{Q}_{\mathcal{G},k}}, [r_\rho]_{\rho\in\mathcal{Q}_{\mathcal{R},k}}, \notag \\
&[r_{\scriptscriptstyle \mathcal{T}_m,s_{\mathcal{T}_m}}]_{m\in\mathcal{Q}_{\mathcal{T},k}}, [r^{\scriptscriptstyle O}_{x_g}]_{g\in\mathcal{Q}_{\mathcal{G},k}}, [r^{\scriptscriptstyle O}_{y_\rho}]_{\rho\in\mathcal{Q}_{R,k}} \Big) = \top \label{eq:fit in goal regions}
\end{align}
$\forall k\in\mathcal{K}$, where we define the sets: $\mathcal{Q}_{\mathcal{Z},k} \coloneqq \{z\in\mathcal{Z}: k'_z = k \}, \mathcal{Q}_{\mathcal{G},k} \coloneqq \{g\in\mathcal{G}: k_g = k \}, \mathcal{Q}_{\mathcal{R},k} \coloneqq \{\rho\in\mathcal{R}: k_r = k \}, \mathcal{Q}_{\mathcal{T},k} \coloneqq \{m\in\{1,\dots,\bar{T}\}: k'_{\mathcal{T}_m} = k \}$, that correspond to the indices of the agents and objects that are in region $k\in\mathcal{K}$.

\begin{example}
As an example, consider $N = 6$ agents, $\mathcal{N} = \{1,\dots,6\}$, $M = 3$ objects, $\mathcal{M} = \{1,2,3\}$ in a workspace that contains $K = 4$ regions of interest, $\mathcal{K} = \{1,\dots,4\}$. Let $t_0 = 0$ and, according to Problem \ref{problem}, take $\text{init}(1) = \text{init}(5) = 1, \text{init}(2) = 2, \text{init}(3) = \text{init}(4) = 3$, and $\text{init}(6) = 4$, i.e., agents $1$ and $5$ are in region $\pi_{\text{init}(1)}=\pi_{\text{init}(5)} = \pi_1$, agent $2$ is in region $\pi_{\text{init}(2)} = \pi_2$, agents $3$ and $4$ are in region $\pi_{\text{init}(3)} = \pi_{\text{init}(4)}=\pi_3$ and agent $6$ is in region $\pi_{\text{init}(6)} = \pi_4$. We also consider $\text{init}_{\scriptscriptstyle O}(1) = 1, \text{init}_{\scriptscriptstyle O}(2) = 2, \text{init}_{\scriptscriptstyle O}(3) = 3$ implying that the $3$ objects are in regions $\pi_1,\pi_2$ and $\pi_3$, respectively. We assume that agents $1,5$ grasp objet $1$, and agents $3,4$ grasp object $3$.
Hence, $\mathcal{AG}_{1,1}(\boldsymbol{q}_1(0),\boldsymbol{x}^{\scriptscriptstyle O}_1(0)) = \mathcal{AG}_{5,1}(\boldsymbol{q}_5(0),\boldsymbol{x}^{\scriptscriptstyle O}_1(0)) = \mathcal{AG}_{3,3}(\boldsymbol{q}_3(0),\boldsymbol{x}^{\scriptscriptstyle O}_3(0)) = \mathcal{AG}_{4,3}(\boldsymbol{q}_4(0),\boldsymbol{x}^{\scriptscriptstyle O}_4(0)) = \mathcal{AG}_{2,0}(\boldsymbol{q}_2(0),\boldsymbol{x}^{\scriptscriptstyle O}(0)) = \mathcal{AG}_{6,0}(\boldsymbol{q}_6(0),\boldsymbol{x}^{\scriptscriptstyle O}(0)) = \top$. Agents $1$ and $5$ aim to cooperatively transport object $1$ to $\pi_4$, agent $2$ aims to grasp object $2$, agents $3$ and $4$ aim to cooperatively transport object $3$ to $\pi_1$ and agent $6$ aims to perform a transition to region $\pi_2$. Therefore, $\mathcal{Z} = \{6\}, \bar{T} =2, \mathcal{T}_1 = \{1,5\}, \mathcal{T}_2 = \{3,4\}$, $\mathcal{T} = \mathcal{T}_1\cup\mathcal{T}_2 = \{1,5,4,3\}$, $\mathcal{G} = \{2\}, \mathcal{R} = \emptyset$, $s_{\mathcal{T}_1} = 1$, $s_{\mathcal{T}_2} = 2$, $\mathcal{S} = \{s_{\mathcal{T}_1},s_{\mathcal{T}_2}\} = \{1,2\}$, $\mathcal{X} = \{x_2\} = \{2\}, \mathcal{Y} = \emptyset$. Moreover, the region indices $k_z,k_\tau,k_g,k_r,k_{\mathcal{T}_m}, k'_z, k'_{\mathcal{T}_m}, z\in\mathcal{Z}=\{6\},\tau\in\mathcal{T}=\{1,5,4,3\}, g\in\mathcal{G}=\{2\},r\in\mathcal{R}=\emptyset,m\in\{1,2\}$, take the form $k_6 = 4, k_1=k_5 = 1, k_2 = 2, k_3=k_4 = 3, k_{\mathcal{T}_1} = 1, k_{\mathcal{T}_3} = 3, k'_6 = 2, k'_{\mathcal{T}_1} = 4, k'_{\mathcal{T}_2} = 1$. Finally, the actions that need to be performed by the agents are $\pi_1 \xrightarrow{T}_{\mathcal{T}_{1,1}} \pi_4$, $2 \xrightarrow{g} 2$, $\pi_3 \xrightarrow{T}_{\mathcal{T}_{2,3}} \pi_1$ and $\pi_4 \rightarrow \pi_2$.
\end{example} 
Next, for each region $\pi_k$, we compute from $\boldsymbol{\mathcal{P}}_s$ a set of configurations for the agents and objects in this region. More specifically, 
\begin{align*}
&([\boldsymbol{q}^\star_z]_{z\in\mathcal{Q}_{\mathcal{Z},k}}, [\boldsymbol{q}^\star_g]_{g\in\mathcal{Q}_{\mathcal{G},k}}, [\boldsymbol{q}^\star_\rho]_{\rho\in\mathcal{Q}_{\mathcal{R},k}}, [\boldsymbol{x}^{\scriptscriptstyle O\star}_{s_{\mathcal{T}_m}}]_{m\in\mathcal{Q}_{\mathcal{T},k}}, \notag \\ 
&[\boldsymbol{x}^{\scriptscriptstyle O\star}_{x_g}]_{g\in\mathcal{Q}_{\mathcal{G},k}},     [\boldsymbol{x}^{\scriptscriptstyle O\star}_{y_\rho}]_{\rho\in\mathcal{Q}_{\mathcal{R},k}}) = \notag \\ 
&\boldsymbol{\mathcal{P}}_{s,1}\Big(r_{\pi_k}, [r_z]_{z\in\mathcal{Q}_{\mathcal{Z},k}}, [r_g]_{g\in\mathcal{Q}_{\mathcal{G},k}}, [r_\rho]_{\rho\in\mathcal{Q}_{\mathcal{R},k}}, \notag \\ 
&[r_{\scriptscriptstyle \mathcal{T}_m,s_{\mathcal{T}_m}}]_{m\in\mathcal{Q}_{\mathcal{T},k}}, [r^{\scriptscriptstyle O}_{x_g}]_{g\in\mathcal{Q}_{\mathcal{G},k}},[r^{\scriptscriptstyle O}_{y_\rho}]_{\rho\in\mathcal{Q}_{R,k}} \Big),
\end{align*}
where we have used the notation of \eqref{eq:fit in goal regions}. Hence, we now have the goal configurations for the agents $\mathcal{Z}$ performing the transitions as well as agents $\mathcal{T}$ performing the cooperative transportations. 

Following Section \ref{subsec:MAS NF}, we define the error functions  $\gamma_{z}:\mathbb{R}^{n_\mathcal{Z}}\rightarrow \mathbb{R}_{\geq 0}$ with $\gamma_{z}(\boldsymbol{q}_z) \coloneqq \lVert \boldsymbol{q}_z - \boldsymbol{q}^\star_z \rVert^2$, $\forall z\in\mathcal{Z}$, $n_\mathcal{Z}\coloneqq \sum_{z\in\mathcal{Z}}n_z$, and $\gamma_{\scriptscriptstyle \mathcal{T}_m}:\mathbb{M}\rightarrow \mathbb{R}_{\geq 0}$ as $\gamma_{\scriptscriptstyle \mathcal{T}_m}(\boldsymbol{x}^{\scriptscriptstyle O}_{s_{\mathcal{T}_m}}) \coloneqq  \lVert \boldsymbol{p}^{\scriptscriptstyle O}_{s_{\mathcal{T}_m}} - \boldsymbol{p}^{\scriptscriptstyle O^\star}_{s_{\mathcal{T}_m}}\rVert^2$, where $\boldsymbol{p}^{\scriptscriptstyle O^\star}_{s_{\mathcal{T}_m}}$ is the position part of $\boldsymbol{x}^{\scriptscriptstyle O^\star}_{s_{\mathcal{T}_m}}$.  

Regarding the collision avoidance, we have the following collision functions: 
\small
\begin{align*}
&\beta_{i,l}(\boldsymbol{q}_i,\boldsymbol{q}_l) \coloneqq  \|\boldsymbol{c}_i(\boldsymbol{q}_i) - \boldsymbol{c}_l(\boldsymbol{q}_l) \|^2 - (r_i + r_l)^2, 
\forall i,l\in\mathcal{N}\backslash\mathcal{T}, i\neq l, \notag \\
&\beta_{i,\scriptscriptstyle O_j}(\boldsymbol{q}_i) \coloneqq  \|\boldsymbol{c}_i(\boldsymbol{q}_i) - \boldsymbol{p}^{\scriptscriptstyle O}_j\|^2 - (r_i + r^{\scriptscriptstyle O}_j)^2, \forall i\in\mathcal{N}\backslash\mathcal{T}, j \in \mathcal{M}\backslash\mathcal{S} \notag \\
&\beta_{i,\scriptscriptstyle \mathcal{T}_m}(\boldsymbol{q}_i,\boldsymbol{x}^{\scriptscriptstyle O}_{s_{\mathcal{T}_m}}) \coloneqq \|\boldsymbol{c}_i(\boldsymbol{q}_i) - \boldsymbol{p}^{\scriptscriptstyle O}_{s_{\mathcal{T}_m}} \|^2 - (r_i + r_{\scriptscriptstyle \mathcal{T}_m, s_{\mathcal{T}_m}})^2, \notag \\ 
&\hspace{25mm}\forall i\in\mathcal{N}\backslash\mathcal{T}, m\in\{1,\dots,\bar{T}\}, \notag \\
&\beta_{\scriptscriptstyle \mathcal{T}_m, \mathcal{T}_\ell }(\boldsymbol{x}^{\scriptscriptstyle O}_{s_{\mathcal{T}_m}}, \boldsymbol{x}^{\scriptscriptstyle O}_{s_{\mathcal{T}_\ell}}) \coloneqq \|\boldsymbol{p}^{\scriptscriptstyle O}_{s_{\mathcal{T}_m}} - \boldsymbol{p}^{\scriptscriptstyle O}_{s_{\mathcal{T}_\ell}} \|^2 - (r_{\scriptscriptstyle \mathcal{T}_m, s_{\mathcal{T}_m}} + r_{\scriptscriptstyle \mathcal{T}_\ell, s_{\mathcal{T}_\ell}})^2, \notag \\
&\hspace{25mm}\forall m,\ell\in\{1,\dots,\bar{T}\}, m \neq \ell, \notag \\
&\beta_{\scriptscriptstyle \mathcal{T}_m, O_j}(\boldsymbol{x}^{\scriptscriptstyle O}_{s_{\mathcal{T}_m}}) \coloneqq \|\boldsymbol{p}^{\scriptscriptstyle O}_{s_{\mathcal{T}_m}} - \boldsymbol{p}^{\scriptscriptstyle O}_j \|^2 - (r_{\scriptscriptstyle \mathcal{T}_m, s_{\mathcal{T}_m}} + r^{\scriptscriptstyle O}_j)^2, \notag \\
&\hspace{25mm}\forall m\in\{1,\dots,\bar{T}\}, j\in\mathcal{M}\backslash\mathcal{S}, \notag \\
& \beta_{i,\pi_k}(\boldsymbol{q}_i) \coloneqq \| \boldsymbol{c}_i(\boldsymbol{q}_i) - \boldsymbol{p}_{\pi_{k} } \|^2 - (r_i + r_{\pi_k})^2, \notag \\
&\hspace{25mm}\forall i\in\mathcal{Z},k\in\mathcal{K}\backslash\{k_z,k'_z\}, \notag \\
&\beta_{\scriptscriptstyle \mathcal{T}_m,\pi_k}(\boldsymbol{x}^{\scriptscriptstyle O}_{s_{\mathcal{T}_m}}) \coloneqq \| \boldsymbol{p}^{\scriptscriptstyle O}_{s_{\mathcal{T}_m}} - \boldsymbol{p}_{\pi_{k} } \|^2 - (r_{\scriptscriptstyle \mathcal{T}_m, s_{\mathcal{T}_m}}  + r_{\pi_k})^2, \notag \\
&\hspace{25mm} \forall m\in\{1,\dots,\bar{T}\}, k\in\mathcal{K}\backslash\{k_{\mathcal{T}_m},k'_{\mathcal{T}_m}\}, \notag \\
& \beta_{i,\scriptscriptstyle \mathcal{W}}(\boldsymbol{q}_i) \coloneqq (r_0 - r_i)^2 - \|\boldsymbol{c}_i(\boldsymbol{q}_i) \|^2, \forall i\in\mathcal{N}\backslash\mathcal{T} \notag \\
& \beta_{\scriptscriptstyle \mathcal{T}_m,\mathcal{W}}(\boldsymbol{x}^{\scriptscriptstyle O}_{s_{\mathcal{T}_m}}) \coloneqq (r_0 - r_{\scriptscriptstyle \mathcal{T}_m, s_{\mathcal{T}_m}})^2 - \|\boldsymbol{p}^{\scriptscriptstyle O}_{s_{\mathcal{T}_m}} \|^2, \forall m\in\{1,\dots,\bar{T}\},
\end{align*}
\normalsize
that incorporate collisions among the navigating agents, the navigating agents and the objects, the transportation agents, the transportation agents and the objects, the navigating agents and the undesired regions, the transportation agents and the undesired regions, the navigating agents and the workspace boundary, and the transportation agents and the workspace boundary, respectively.
Therefore, by following the procedure described in Section \ref{subsec:MAS NF}, we can form the total obstacle function $G:\mathbb{R}^{n_\mathcal{Z}}\times\mathbb{M}^{|\mathcal{S}| }  \to\mathbb{R}_{\geq 0}$ and thus, define the navigation function \cite{Koditchek92,Loizou2006} $\varphi:\mathbb{R}^{n_\mathcal{Z}}\times\mathbb{M}^{|\mathcal{S}| }   \to [0,1]$ as 
\begin{equation*}
\varphi(\boldsymbol{q}_{\scriptscriptstyle \mathcal{Z}},\boldsymbol{x}^{\scriptscriptstyle O}_{\scriptscriptstyle \mathcal{S}}) \coloneqq \frac{ \gamma(\boldsymbol{q}_{\scriptscriptstyle \mathcal{Z}},\boldsymbol{x}^{\scriptscriptstyle O}_{\scriptscriptstyle \mathcal{S}}) }{ \Big( [\gamma(\boldsymbol{q}_{\scriptscriptstyle \mathcal{Z}},\boldsymbol{x}^{\scriptscriptstyle O}_{\scriptscriptstyle \mathcal{S}})]^\kappa + G(\boldsymbol{q}_{\scriptscriptstyle \mathcal{Z}},\boldsymbol{x}^{\scriptscriptstyle O}_{\scriptscriptstyle \mathcal{S}}) \Big)^{\frac{1}{\kappa}}},
\end{equation*}
where $\boldsymbol{x}^{\scriptscriptstyle O}_{\scriptscriptstyle \mathcal{S}} \coloneqq [ (\boldsymbol{x}^{\scriptscriptstyle O}_{s_{\scriptscriptstyle \mathcal{T}_m}} )^\top ]^\top_{m\in\{1,\dots,\bar{T}\}} \in \mathbb{M}^{|\mathcal{S}|}$, $\gamma(\boldsymbol{q}_{\scriptscriptstyle \mathcal{Z}},\boldsymbol{x}^{\scriptscriptstyle O}_{\scriptscriptstyle \mathcal{S}}) \coloneqq \sum_{z\in\mathcal{Z}} \gamma_z(\boldsymbol{q}_z) + \sum_{m\in\{1,\dots,\bar{T}\}} \gamma_{\scriptscriptstyle \mathcal{T}_m}(\boldsymbol{x}^{\scriptscriptstyle O}_{\scriptscriptstyle s_{\mathcal{T}_m}})$ and $\kappa > 0$ is a positive gain used to derive the proof correctness of $\varphi$ \cite{Koditchek92,Loizou2006}.  Note that, a sufficient condition for avoidance of the undesired regions and avoidance of collisions and singularities is $\varphi(\boldsymbol{q}_{\scriptscriptstyle \mathcal{Z}},\boldsymbol{x}^{\scriptscriptstyle O}_{\scriptscriptstyle \mathcal{S}}) < 1$.

Next, we design the feedback control protocols $\boldsymbol{u}_z: \mathbb{R}^{n_z}\times\mathbb{R}^{n_z}\to \mathbb{R}^6, \boldsymbol{u}_\tau:\mathbb{S}_\tau\times \mathbb{S}^{\scriptscriptstyle O}_{s_{T_m}}\times\mathbb{R}^6$, $\forall z\in\mathcal{Z}, \tau\in\mathcal{T}_m, m\in\{1,\dots,\bar{T}\}$ as follows:
\begin{subequations} \label{eq:control_protocol}
\begin{align}
&\boldsymbol{u}_z(\boldsymbol{q}_z,\boldsymbol{x}^{\scriptscriptstyle O}_{\scriptscriptstyle \mathcal{S}},\dot{\boldsymbol{q}}_z)  = \boldsymbol{g}_{q_z}(\boldsymbol{q}_z) -  \nabla_{\boldsymbol{q}_z}\varphi(\boldsymbol{q}_{\scriptscriptstyle \mathcal{Z}},\boldsymbol{x}^{\scriptscriptstyle O}_{\scriptscriptstyle \mathcal{S}}) - \boldsymbol{K}_z\dot{\boldsymbol{q}}_z, 	\label{eq:control_protocol_transition} \\
&\boldsymbol{u}_\tau(\boldsymbol{q}_\tau,\boldsymbol{x}^{\scriptscriptstyle O}_{s_{\mathcal{T}_m}}, \boldsymbol{v}^{\scriptscriptstyle O}_{s_{\mathcal{T}_m}} ) =   [\boldsymbol{J}^{\scriptscriptstyle O}_{\tau,s_{\mathcal{T}_m}}(\boldsymbol{q}_\tau)]^{-\top} \Big\{ c_\tau \Big(\boldsymbol{g}_{\scriptscriptstyle O}(\boldsymbol{x}^{\scriptscriptstyle O}_{s_{\mathcal{T}_m}}) - \notag \\
&  [\boldsymbol{J}^{\scriptscriptstyle O}_{s_{\mathcal{T}_m}}(\boldsymbol{x}^{\scriptscriptstyle O}_{s_{\mathcal{T}_m}})]^\top \nabla_{\scriptscriptstyle \boldsymbol{x}^{\scriptscriptstyle O}_{s_{\mathcal{T}_m}}} \varphi(\boldsymbol{q}_{\scriptscriptstyle \mathcal{Z}},\boldsymbol{x}^{\scriptscriptstyle O}_{\scriptscriptstyle \mathcal{S}}) - \boldsymbol{v}^{\scriptscriptstyle O}_{s_{\mathcal{T}_m}} \Big) \Big\} + \boldsymbol{g}_{\tau}(\boldsymbol{q}_\tau),  	\label{eq:control_protocol_transportation}
\end{align}
\end{subequations}
where $c_\tau$ are load sharing coefficients, with the properties $c_\tau > 0$, $\forall \tau\in\mathcal{T}_m$, $\sum_{\tau\in\mathcal{T}_m}c_\tau = 1$, $\forall m\in\{1,\dots,\bar{T}\}$, $\boldsymbol{K}_z = \text{diag}\{k_z\}\in\mathbb{R}^{n_z \times n_z}$, with $k_z > 0, \forall z\in\mathcal{Z}$, is a constant positive definite gain matrix. 
The proof of convergence of the closed loop system is stated in the next Lemma. 

\begin{lemma} \label{lem:agent transition}
Consider the sets of agent $\mathcal{Z}, \mathcal{T}, \mathcal{G}, \mathcal{R}$ and the set of objects $\mathcal{S},\mathcal{X},\mathcal{R}$ in their respective regions interest, as defined above, described by the dynamics \eqref{eq:manipulator dynamics}, \eqref{eq:object dynamics}, \eqref{eq:coupled dynamics} at $t_0 >0$. Then, under the assumptions that: (i) the actions $g\xrightarrow{g}x_g, \rho\xrightarrow{r}y_\rho$ are guaranteed, (ii) \eqref{eq:fit in goal regions} holds and (iii) the robots and objects operate in singularity-free (kinematic- and representation ones, respectively) configurations, the control protocols \eqref{eq:control_protocol} guarantee the existence of a $t_f > t_0$ such that $\pi_{k_z}\rightarrow_z\pi_{k'_z}$ and $\pi_{k_{\scriptscriptstyle \mathcal{T}_m}}\xrightarrow{T}_{\mathcal{T}_m,s_{\mathcal{T}_m}} \pi_{k'_{\scriptscriptstyle \mathcal{T}_m}}, \forall z\in\mathcal{Z}, m\in\{1,\dots,\bar{T}\}$, according to Def. \ref{def:agent transition} and \ref{def:agent-object transition}, respectively.
\end{lemma}
\begin{IEEEproof}
Define $\bar{\mathcal{T}}\coloneqq \{1,\dots,\bar{T}\}$ and, following the notation of Section \ref{sec:Model and PF}, consider the stacked vector states
$\boldsymbol{x}_{\scriptscriptstyle \mathcal{T}_m,s_{\mathcal{T}_m}}\coloneqq [\boldsymbol{q}^\top_{\mathcal{T}_m}, \dot{\boldsymbol{q}}^\top_{\mathcal{T}_m}, (\boldsymbol{x}^{\scriptscriptstyle O}_{s_{\mathcal{T}_m}})^\top, (\boldsymbol{v}^{\scriptscriptstyle O}_{s_{\mathcal{T}_m}})^\top]^\top, m\in\bar{\mathcal{T}}$,
$\boldsymbol{d} \coloneqq [\boldsymbol{q}^\top_{\scriptscriptstyle \mathcal{Z}},\dot{\boldsymbol{q}}^\top_{\scriptscriptstyle \mathcal{Z}},[\boldsymbol{x}^\top_{\scriptscriptstyle \mathcal{T}_m,s_{\mathcal{T}_m}}]^\top_{m\in\bar{\mathcal{T}}}]^\top$
as well as the domain: 
$\mathbb{D} \coloneqq \mathbb{R}^{n_\mathcal{Z}}\times \mathbb{R}^{n_\mathcal{Z}}\times\mathbb{S}_{\mathcal{T}_1}\times\dots\times\mathbb{S}_{\mathcal{T}_{\bar{T}}} \times \mathbb{R}^{n_{\mathcal{T}}} \times \mathbb{S}^{\scriptscriptstyle O}_{s_{\mathcal{T}_1}}\times\dots\times \mathbb{S}^{\scriptscriptstyle O}_{s_{\mathcal{T}_{\bar{T}}}}\times \mathbb{R}^{6|\mathcal{S}|},$
where $\mathbb{S}_{\mathcal{T}_m} \coloneqq \prod_{\tau\in\mathcal{T}_m} \mathbb{S}_{\tau}, \forall m\in\bar{\mathcal{T}}$, and $n_{\mathcal{T}} \coloneqq \sum_{m\in\bar{\mathcal{T}}}\sum_{\tau\in\mathcal{T}_m}n_\tau$.
 Consider now the candidate Lyapunov function $V:\mathbb{D}\rightarrow \mathbb{R}_{\geq 0}$, with
\begin{align*}
V(\boldsymbol{d}) =&
\varphi(\boldsymbol{q}_{\scriptscriptstyle \mathcal{Z}},\boldsymbol{x}^{\scriptscriptstyle O}_{\scriptscriptstyle \mathcal{S}}) + \frac{1}{2}\sum_{z\in\mathcal{Z}}\dot{\boldsymbol{q}}^T_z  \boldsymbol{B}_z(\boldsymbol{q}_z) \dot{\boldsymbol{q}}_z  + \notag \\
&\frac{1}{2}\sum_{m\in \bar{\mathcal{T}}}[\boldsymbol{v}^{\scriptscriptstyle O}_{s_{\mathcal{T}_m}} ]^\top \widetilde{\boldsymbol{M}}_{\scriptscriptstyle \mathcal{T}_m, s_{\mathcal{T}_m}}(\boldsymbol{x}_{\scriptscriptstyle \mathcal{T}_m,s_{\mathcal{T}_m}})\boldsymbol{v}^{\scriptscriptstyle O}_{s_{\mathcal{T}_m}}.
\end{align*}
Note that, since no collisions occur and the robots and objects have zero velocity at $t_0$, we conclude that $V_0 \coloneqq V(\boldsymbol{d}(t_0)) =  \varphi(\boldsymbol{q}_{\scriptscriptstyle \mathcal{Z}}(t_0),\boldsymbol{x}^{\scriptscriptstyle O}_{\scriptscriptstyle \mathcal{S}}(t_0)) =: \varphi_0 < 1$, and hence $\boldsymbol{d}(t_0)\in\widetilde{\mathbb{D}} \coloneqq \{\boldsymbol{d}\in\mathbb{D} : \varphi(\boldsymbol{q}_{\scriptscriptstyle \mathcal{Z}},\boldsymbol{x}^{\scriptscriptstyle O}_{\scriptscriptstyle \mathcal{S}}) \leq \varphi_0 < 1\}$.  By considering the closed loop system $\tfrac{\partial}{\partial t}\boldsymbol{d} = \boldsymbol{f}_{\text{cl}}(\boldsymbol{d})$ (An explicit expression for $\boldsymbol{f}_{\text{cl}}$ can be obtained by combining \eqref{eq:manipulator dynamics}, \eqref{eq:coupled dynamics}, \eqref{eq:control_protocol}), we can verify the locally Lipschitz property of $\boldsymbol{f}_{\text{cl}}$, and thus the existence of a unique maximal solution $\boldsymbol{d}:[t_0,t_{\max}) \to \widetilde{\mathbb{D}}$, for a finite time instant $t_{\max} > t_0$. 
By differentiating $V$ and substituting \eqref{eq:manipulator dynamics}, \eqref{eq:coupled dynamics}, we obtain 
\small
\begin{align}
&\dot{V} =  \notag\sum\limits_{z\in\mathcal{Z}}\Big\{ [\nabla_{\boldsymbol{q}_z}\varphi(\boldsymbol{q}_{\scriptscriptstyle \mathcal{Z}},\boldsymbol{x}^{\scriptscriptstyle O}_{\scriptscriptstyle \mathcal{S}})]^\top \dot{\boldsymbol{q}}_z + \dot{\boldsymbol{q}}^\top_z\Big( \boldsymbol{\tau}_z - \boldsymbol{N}_z(\boldsymbol{q}_z,\dot{\boldsymbol{q}}_z)\dot{\boldsymbol{q}}_z -
\notag \\
&\boldsymbol{g}_{q_z}(\boldsymbol{q}_z) \Big) +  \frac{1}{2}\dot{\boldsymbol{q}}_z^\top\dot{\boldsymbol{M}}_z(\boldsymbol{q}_z) \dot{\boldsymbol{q}}_z \Big\} + \sum\limits_{m\in\bar{\mathcal{T}}} \Big\{ [\nabla_{\boldsymbol{x}^{\scriptscriptstyle O}_{\scriptscriptstyle s_{\mathcal{T}_m}}} \varphi(\boldsymbol{q}_{\scriptscriptstyle \mathcal{Z}},\boldsymbol{x}^{\scriptscriptstyle O}_{\scriptscriptstyle \mathcal{S}}) ]^\top  \dot{\boldsymbol{x}}^{\scriptscriptstyle O}_{\scriptscriptstyle s_{\mathcal{T}_m}} \notag \\
& +   [\boldsymbol{v}^{\scriptscriptstyle O}_{\scriptscriptstyle s_{\mathcal{T}_m}}]^\top \Big( \sum\limits_{\tau\in\mathcal{T}_m} [\boldsymbol{J}^{\scriptscriptstyle O}_{\tau,\scriptscriptstyle s_{\mathcal{T}_m}}(\boldsymbol{q}_\tau)]^\top \boldsymbol{u}_\tau  - \widetilde{\boldsymbol{C}}_{\scriptscriptstyle \mathcal{T}_m, s_{\mathcal{T}_m}}(\boldsymbol{x}_{\scriptscriptstyle \mathcal{T}_m,s_{\mathcal{T}_m}}) \boldsymbol{v}^{\scriptscriptstyle O}_{\scriptscriptstyle s_{\mathcal{T}_m}}  \notag \\
&\hspace{5mm}- \sum\limits_{\tau\in\mathcal{T}_m} [\boldsymbol{J}^{\scriptscriptstyle O}_{\scriptscriptstyle \tau,s_{\mathcal{T}_m}}(\boldsymbol{q}_\tau)]^\top \boldsymbol{g}_\tau(\boldsymbol{q}_\tau) - \boldsymbol{g}_{\scriptscriptstyle O}(\boldsymbol{x}^{\scriptscriptstyle O}_{\scriptscriptstyle s_{\mathcal{T}_m}}) 
\Big) + \notag\\
&\hspace{5mm}  \frac{1}{2} [\boldsymbol{v}^{\scriptscriptstyle O}_{s_{\mathcal{T}_m}}]^\top \dot{\widetilde{\boldsymbol{M}}}_{\scriptscriptstyle \mathcal{T}_m,s_{\mathcal{T}_m}}(\boldsymbol{x}_{\scriptscriptstyle \mathcal{T}_m,s_{\mathcal{T}_m}}) \boldsymbol{v}^{\scriptscriptstyle O}_{\scriptscriptstyle s_{\mathcal{T}_m}} \Big\}, \notag
\end{align}
\normalsize
$\forall \boldsymbol{d}\in\widetilde{\mathbb{D}}$, where we have also used the fact that $\boldsymbol{f}_z = 0,\forall z\in\mathcal{Z}$, since the agents performing transportation actions are not in contact with any objects (and there are no collisions in $\widetilde{\mathbb{D}}$).   
By employing Lemma \ref{lem:coupled dynamics skew symmetry} as well as \eqref{eq:object dynamics 1}, $\dot{V}$ becomes:
\small
\begin{align}
&\dot{V} =  \notag\sum\limits_{z\in\mathcal{Z}}\dot{\boldsymbol{q}}^\top_z\Big(\nabla_{\boldsymbol{q}_z}\varphi(\boldsymbol{q}_{\scriptscriptstyle \mathcal{Z}},\boldsymbol{x}^{\scriptscriptstyle O}_{\scriptscriptstyle \mathcal{S}}) + \boldsymbol{\tau}_z - \boldsymbol{g}_{q_z}(\boldsymbol{q}_z) \Big)     + \notag \\
&\sum\limits_{m\in\bar{\mathcal{T}}} [\boldsymbol{v}^{\scriptscriptstyle O}_{\scriptscriptstyle s_{\mathcal{T}_m}}]^\top \Big( [\boldsymbol{J}^{\scriptscriptstyle O}_{\scriptscriptstyle s_{\mathcal{T}_m}}(\boldsymbol{x}^{\scriptscriptstyle O}_{\scriptscriptstyle s_{\mathcal{T}_m}})]^\top \nabla_{\boldsymbol{x}^{\scriptscriptstyle O}_{\scriptscriptstyle s_{\mathcal{T}_m}}} \varphi(\boldsymbol{q}_{\scriptscriptstyle \mathcal{Z}},\boldsymbol{x}^{\scriptscriptstyle O}_{\scriptscriptstyle \mathcal{S}}) + \notag \\
& \sum\limits_{\tau\in\mathcal{T}_m} [\boldsymbol{J}^{\scriptscriptstyle O}_{\tau,\scriptscriptstyle s_{\mathcal{T}_m}}(\boldsymbol{q}_\tau)]^\top \boldsymbol{u}_\tau  -  \sum\limits_{\tau\in\mathcal{T}_m}  [\boldsymbol{J}^{\scriptscriptstyle O}_{\tau,\scriptscriptstyle s_{\mathcal{T}_m}}(\boldsymbol{q}_\tau)]^\top \boldsymbol{g}_\tau(\boldsymbol{q}_\tau) - \notag\\
& \boldsymbol{g}_{\scriptscriptstyle O}(\boldsymbol{x}^{\scriptscriptstyle O}_{\scriptscriptstyle s_{\mathcal{T}_m}}) 
\Big), \notag
\end{align}
\normalsize
and after substituting \eqref{eq:control_protocol}:
$\dot{V} = -\sum_{z\in\mathcal{Z}}\dot{\boldsymbol{q}}_z\boldsymbol{K}_z\dot{\boldsymbol{q}}_z - \sum_{m\in\widetilde{\mathcal{T}}}\| \boldsymbol{v}^{\scriptscriptstyle O}_{\scriptscriptstyle s_{\mathcal{T}_m}} \|^2$,
$\forall \boldsymbol{d}\in\widetilde{\mathbb{D}}$, which is strictly negative unless $\dot{\boldsymbol{q}}_z = \boldsymbol{0}$, $\boldsymbol{v}^{\scriptscriptstyle O}_{\scriptscriptstyle s_{\mathcal{T}_m}} = \boldsymbol{0}, \forall z\in\mathcal{Z},m\in\widetilde{\mathcal{T}}$. Since $\boldsymbol{J}^{\scriptscriptstyle O}_{\scriptscriptstyle \tau,s_{\mathcal{T}_m}}(\boldsymbol{q}_{\tau})$ is always non-singular, and $\boldsymbol{J}_\tau(\boldsymbol{q}_\tau(t))$ has full-rank by assumption for the maximal solution, $\forall \tau\in\mathcal{T}_m,m\in\widetilde{\mathcal{T}}$, the latter implies also that $\dot{\boldsymbol{q}}_\tau = \boldsymbol{0}$, $\forall \tau\in\mathcal{T}_m,m\in\widetilde{\mathcal{T}}$. Hence, $V(\boldsymbol{d}(t)) \leq V_0 < 1$, $\forall t\in[t_0,t_{\max})$, which suggests that $\varphi(\boldsymbol{q}_{\scriptscriptstyle \mathcal{Z}}(t),\boldsymbol{x}^{\scriptscriptstyle O}_{\scriptscriptstyle \mathcal{S}}(t)) \leq \varphi_0 < 1$ and $\boldsymbol{d}(t)\in\widetilde{\mathbb{D}}$, $\forall t\in[t_0,t_{\max})$. Therefore, since $\widetilde{\mathbb{D}}$ is compact, the solution $\boldsymbol{d}(t)$ is defined over the entire time horizon in $\widetilde{\mathbb{D}}$ \cite{Khalil}, i.e. $\boldsymbol{d}:[t_0,\infty)\to\widetilde{\mathbb{D}}$. Moreover, according to La Salle's Invariance Principle \cite{Khalil}, the system will converge to the largest invariant set contained in the set $\{\boldsymbol{d}\in\widetilde{\mathbb{D}} : \dot{\boldsymbol{q}}_z = \boldsymbol{0}, \boldsymbol{v}^{\scriptscriptstyle O}_{\scriptscriptstyle s_{\mathcal{T}_m}}=\boldsymbol{0}, \forall z\in\mathcal{Z}, m\in\widetilde{\mathcal{T}} \}$. In order for this set to be invariant, we require that $\ddot{{\boldsymbol{q}}}_z = \boldsymbol{0}$, $\dot{\boldsymbol{v}}^{\scriptscriptstyle O}_{\scriptscriptstyle s_{\mathcal{T}_m}} = \boldsymbol{0}$, which, by employing \eqref{eq:control_protocol}, \eqref{eq:manipulator dynamics}, \eqref{eq:coupled dynamics}, and the assumption of non-singular $\boldsymbol{J}^{\scriptscriptstyle O}_{\scriptscriptstyle s_{\mathcal{T}_m}}(\boldsymbol{x}^{\scriptscriptstyle O}_{\scriptscriptstyle s_{\mathcal{T}_m}}(t))$, $\forall t\in\mathbb{R}_{\geq 0}$, implies that $\nabla_{\boldsymbol{q}_z}\varphi(\boldsymbol{q}_{\scriptscriptstyle \mathcal{Z}},\boldsymbol{x}^{\scriptscriptstyle O}_{\scriptscriptstyle \mathcal{S}}) = \boldsymbol{0}$, $\nabla_{\boldsymbol{x}^{\scriptscriptstyle O}_{\scriptscriptstyle s_{\mathcal{T}_m}}} \varphi(\boldsymbol{q}_{\scriptscriptstyle \mathcal{Z}},\boldsymbol{x}^{\scriptscriptstyle O}_{\scriptscriptstyle \mathcal{S}}) = \boldsymbol{0}$, $\forall z\in\mathcal{Z}, m\in\widetilde{\mathcal{T}}$. Since $\varphi$ is a navigation function \cite{Loizou2006}, this condition is true only at the destination configurations (i.e., where $\gamma(\boldsymbol{q}_{\scriptscriptstyle \mathcal{Z}},\boldsymbol{x}^{\scriptscriptstyle O}_{\scriptscriptstyle \mathcal{S}}) = 0$) and a set of isolated saddle points. By choosing $\kappa$ sufficiently large, the region of attraction of the saddle points is a set of measure zero \cite{koditschek1989application,koditschek1990robot}. Thus, the system converges to the destination configuration from almost everywhere, i.e., $\lVert \boldsymbol{q}_z(t) - \boldsymbol{q}^\star_z\rVert \rightarrow 0$ and $\lVert \boldsymbol{p}^{\scriptscriptstyle O}_{\scriptscriptstyle s_{\mathcal{T}_m}}(t) - \boldsymbol{p}^{\scriptscriptstyle O^\star}_{\scriptscriptstyle s_{\mathcal{T}_m}}\rVert \rightarrow 0$. Therefore, there exist finite time instants $t_{f_z},t_{f_m} > t_0$, such that $\mathcal{A}_z(\boldsymbol{q}_z(t_{f_z}))\subset \pi_{k'_z}$ and $\mathcal{A}_\tau(\boldsymbol{q}_\tau(t_{f_m})),\mathcal{O}_{\scriptscriptstyle s_{\mathcal{T}_m}}(\boldsymbol{x}^{\scriptscriptstyle O}_{\scriptscriptstyle s_{\mathcal{T}_m}}(t_{f_v}))\subset \pi_{k'_{\mathcal{T}_m}}$, with inter-agent collision avoidance, $\forall z\in\mathcal{Z}, \tau \in\mathcal{T}_m, m\in\widetilde{\mathcal{T}}$. Since the actions $g\xrightarrow{g}x_g$, $\rho \xrightarrow{r} y_\rho$ are also performed, we denote as $t_{f_g}, t_{f_\rho}$ the times that these actions have been completed, $g\in\mathcal{G}, \rho\in\mathcal{R}$. Hence, by setting $t_f \coloneqq \max\{ \max\limits_{z\in\mathcal{Z}}t_{f_z},\max\limits_{m\in\widetilde{\mathcal{T}}}t_{f_m}, \max\limits_{g\in\mathcal{G}}t_{f_g}, \max\limits_{\rho\in\mathcal{R}}t_{f_\rho}\}$, all the actions of all agents will be completed at $t_f$.
\end{IEEEproof}

\begin{remark}
We could modify the dynamic model \eqref{eq:object dynamics} by employing the physical acceleration $\ddot{\boldsymbol{x}}^{\scriptscriptstyle O}_j$ instead of the generalized accelerations $\dot{\boldsymbol{v}}^{\scriptscriptstyle O}_j$, $j\in\mathcal{M}$. In that way, we would avoid using the term $\boldsymbol{J}^{\scriptscriptstyle O}_{j}$ and hence ensure that representation singularities (when $|\eta^{\scriptscriptstyle O}_{j,2} | = \tfrac{\pi}{2}$) do not affect our scheme. Note that the actual difference lies in the use of $\dot{\boldsymbol{\eta}}^{\scriptscriptstyle O}_j$ instead of $\boldsymbol{\omega}^{\scriptscriptstyle O}_j, j\in\mathcal{M}$. Feedback, however, of $\dot{\boldsymbol{\eta}}^{\scriptscriptstyle O}_j$ is not a realistic assumption, since most sensors provide on-line measurements of the angular velocity $\boldsymbol{\omega}^{\scriptscriptstyle O}_j$ and hence, the conversion via $\boldsymbol{J}^{\scriptscriptstyle O}_{j}$ cannot be avoided. 
\end{remark}

\begin{remark}
The fact that we consider fully actuated holonomic mobile bases is not restrictive, since a similar analysis can be performed for non-holonomic agents (see \cite{tanner2001nonholonomic}). Note also that in our analysis we do not take into account potential collisions between agents that grasp and transport the same object, since we just consider the bounded spherical volume of the system. This specification constitutes part of our ongoing work.
\end{remark}

\subsection{High-Level Plan Generation} \label{sec:high level plan}

The second part of the solution is the derivation of a high-level plan that satisfies the given LTL formulas $\phi_i$ and $\phi^{\scriptscriptstyle O}_j$ and can be generated by using standard techniques from automata-based formal verification methodologies. Thanks to (i) the proposed control laws that allow agent transitions and object transportations $\pi_k\rightarrow_i\pi_{k'}$ and $\pi_k\xrightarrow{T}_{\mathcal{T},j}\pi_{k'}$, respectively, and (ii) the off-the-self control laws that guarantee grasp and release actions $i\xrightarrow{g}j$ and $i\xrightarrow{r}j$, we can abstract the behavior of the agents using a finite transition system as presented in the sequel.

\begin{definition} \label{def:TS objects all agents}
The coupled behavior of the overall system of all the $N$ agents and $M$ objects is modeled by the transition system $\mathcal{TS} = (\Pi_s,\Pi^{\text{init}}_s, \rightarrow_{s},\mathcal{AG}, \Psi, \mathcal{L}, \Lambda, \boldsymbol{P}_s,\chi)$,
where 
\begin{enumerate}[label=(\roman*), align=left, leftmargin=0pt, listparindent=\parindent, labelwidth=0pt, itemindent=!]
\item $\Pi_s\subset \bar{\Pi}\times\bar{\Pi}^{\scriptscriptstyle O}\times\bar{\mathcal{AG}}$ is the set of states; 
$\bar{\Pi}\coloneqq\Pi_1\times\cdots\times\Pi_N$ and $\bar{\Pi}^{\scriptscriptstyle O}\coloneqq\Pi^{\scriptscriptstyle O}_1\times\cdots\times\Pi^{\scriptscriptstyle O}_M$ are the set of states-regions that the agents and the objects can be at, with $\Pi_i  = \Pi^{\scriptscriptstyle O}_j = \Pi,\forall i\in\mathcal{N},j\in\mathcal{M}$; 
$\mathcal{AG} \coloneqq \mathcal{AG}_1\times\cdots\times\mathcal{AG}_N$ is the set of boolean grasping variables introduced in Section \ref{sec:Model and PF}, with
$\mathcal{AG}_i \coloneqq \{\mathcal{AG}_{i,0}\}\cup\{[\mathcal{AG}_{i,j}]_{j\in\mathcal{M}}\}, \forall i\in\mathcal{N}$. 
By defining 
$\bar{\pi} \coloneqq \left(\pi_{k_1},\cdots,\pi_{k_N}\right),\bar{\pi}_{\scriptscriptstyle O}  \coloneqq (\pi_{\scriptscriptstyle k^{\scriptscriptstyle O}_1},\cdots,\pi_{\scriptscriptstyle k^{\scriptscriptstyle O}_M}), \bar{w}=\left(w_1,\cdots,w_N\right)$, with $\pi_{k_i},\pi_{k^{\scriptscriptstyle O}_j}\in\Pi$ (i.e., $k_i,k^{\scriptscriptstyle O}_j\in\mathcal{K},\forall i\in\mathcal{N},j\in\mathcal{M}$) and $w_i\in\mathcal{AG}_i, \forall i\in\mathcal{N}$, then the coupled state $\pi_s \coloneqq (\bar{\pi},\bar{\pi}_{\scriptscriptstyle O},\bar{w})$ belongs to $\Pi_s$, i.e., $(\bar{\pi},\bar{\pi}_{\scriptscriptstyle O},\bar{w})\in\Pi_s$ if
\begin{enumerate}
\item $\mathcal{P}_{s,0}\Big(r_{\pi_k}, [r_i]_{i\in\{ i\in\mathcal{N}: k_i = k \} }, [r^{\scriptscriptstyle O}_j]_{j\in\{ j\in\mathcal{M}:k^{\scriptscriptstyle O}_j= k \} }\Big) = \top$, i.e., the respective agents and objects fit in the region, $\forall k\in\mathcal{K}$, 
\item $k_i = k^{\scriptscriptstyle O}_j$ for all $i\in\mathcal{N}, j\in\mathcal{M}$ such that $w_i = \mathcal{AG}_{i,j} = \top$, i.e., an agent must be in the same region with the object it grasps,
\end{enumerate} 
\item $\Pi^{\text{init}}_s\subset\Pi_s$ is the initial set of states at $t=0$, which, owing to \textbf{(i)}, satisfies the conditions of Problem \ref{problem},\\
\item $\rightarrow_s\subset \Pi_s\times\Pi_s$ is a transition relation defined as follows: given the states $\pi_s, \widetilde{\pi}_s\in\Pi$, with
\small
\begin{align}
\pi_s \coloneqq & (\bar{\pi},\bar{\pi}_{\scriptscriptstyle O},\bar{w}) \coloneqq (\pi_{k_1},\dots,\pi_{k_N}, \pi_{k^{\scriptscriptstyle O}_1},\dots,\pi_{k^{\scriptscriptstyle O}_M},w_1,\dots,w_N), \notag \\
\widetilde{\pi}_s \coloneqq & ( \widetilde{\bar{\pi}},\widetilde{\bar{\pi}}_{\scriptscriptstyle O},\widetilde{\bar{w}}) \coloneqq (\pi_{\widetilde{k}_1},\dots,\pi_{\widetilde{k}_N}, \pi_{\widetilde{k}^{\scriptscriptstyle O}_1},\dots,\pi_{\widetilde{k}^{\scriptscriptstyle O}_1},\widetilde{w}_1,\dots,\widetilde{w}_N), \label{eq:pi_s}
\end{align}
\normalsize
a transition $\pi_s \rightarrow_s \widetilde{\pi}_s $ occurs if all the following hold:

\begin{enumerate}
	\item $\nexists i\in\mathcal{N}, j\in\mathcal{M}$ such that $w_i = \mathcal{AG}_{i,j} = \top$, $\widetilde{w}_i = \mathcal{AG}_{i,0} = \top$, (or $w_i = \mathcal{AG}_{i,0} = \top$, $\widetilde{w}_i = \mathcal{AG}_{i,j} = \top$) and $k_i \neq \widetilde{k}_i$, i.e., there are no simultaneous grasp/release and navigation actions,
	\item $\nexists i\in\mathcal{N}, j\in\mathcal{M}$ such that $w_i = \mathcal{AG}_{i,j} = \top$, $\widetilde{w}_i = \mathcal{AG}_{i,0} = \top$, (or $w_i = \mathcal{AG}_{i,0} = \top$, $\widetilde{w}_i = \mathcal{AG}_{i,j} = \top$) and $k_i = k^{\scriptscriptstyle O}_j \neq \widetilde{k}_i=\widetilde{k}^{\scriptscriptstyle O}_j$, i.e., there are no simultaneous grasp/release and transportation actions,
	\item $\nexists i\in\mathcal{N}, j,j'\in\mathcal{M}$, with $j\neq j'$, such that $w_i = \mathcal{AG}_{i,j} = \top$ and $\widetilde{w}_i = \mathcal{AG}_{i,j'} = \top$ ($w_i = \mathcal{AG}_{i,j'} = \top$ and $\widetilde{w}_i = \mathcal{AG}_{i,j'} = \top$), i.e., there are no simultaneous grasp and release actions,
	\item $\nexists j\in\mathcal{M}$ such that $k^{\scriptscriptstyle O}_j \neq \widetilde{k}^{\scriptscriptstyle O}_j$ and $w_i \neq \mathcal{AG}_{i,j}, \forall i\in\mathcal{N}$ ( or $\widetilde{w}_i\neq \mathcal{AG}_{i,j}, \forall i\in\mathcal{N}$), i.e., there is no transportation of a non-grasped object,
	\item $\nexists j\in\mathcal{M},\mathcal{T}\subseteq \mathcal{N}$ such that $k^{\scriptscriptstyle O}_j \neq \widetilde{k}^{\scriptscriptstyle O}_j$ and $\Lambda(m^{\scriptscriptstyle O}_j, \zeta_{\mathcal{T}}) = \bot$, where $w_i = \widetilde{w}_i = \mathcal{AG}_{i,j} = \top \Leftrightarrow i\in\mathcal{T}$, i.e., the agents grasping an object are powerful enough to transfer it,
\end{enumerate}

\item $\Psi \coloneqq \bar{\Psi}\cup\bar{\Psi}^{\scriptscriptstyle O}$ with $\bar{\Psi}=\bigcup_{i\in\mathcal{N}}\Psi_{i}$ and $\bar{\Psi}^{\scriptscriptstyle O} = \bigcup_{j\in\mathcal{M}}\Psi^{\scriptscriptstyle O}_j$, are the atomic propositions of the agents and objects, respectively, as defined in Section \ref{sec:Model and PF}.

\item $\mathcal{L}:\Pi_s \rightarrow 2^\Psi$ is a labeling function defined as follows: Given a state $\pi_s$ as in \eqref{eq:pi_s} and $\psi_s \coloneqq \Big( \bigcup_{i\in\mathcal{N}}\psi_i\Big)\bigcup \Big(\bigcup_{j\in\mathcal{M}}\psi^{\scriptscriptstyle O}_j\Big)$ with $\psi_i\in2^{\Psi_i},\psi^{\scriptscriptstyle O}_j\in2^{\Psi^{\scriptscriptstyle O}_j}$, then $\psi_s\in\mathcal{L}(\pi_s)$ if  $\psi_i\in\mathcal{L}_i(\pi_{k_i})$ and $\psi^{\scriptscriptstyle O}_j\in\mathcal{L}^{\scriptscriptstyle O}_j(\pi_{k^{\scriptscriptstyle O}_j}), \forall i\in\mathcal{N},j\in\mathcal{M}$.

\item $\Lambda$ and $\boldsymbol{P}_s$ as defined in Section \ref{sec:Model and PF}.

\item $\chi: (\to_s) \to \mathbb{R}_{\geq 0}$ is a function that assigns a cost to each transition $\pi_s \to_s \widetilde{\pi}_s$. This cost might be related to the distance of the agents' regions in $\pi_s$ to the ones in $\widetilde{\pi}_s$, combined with the cost efficiency of the agents involved in transport tasks (according to $\zeta_i, i\in\mathcal{N}$).
\end{enumerate} 
\end{definition}


Next, we form the global LTL formula $\phi \coloneqq (\land_{i\in\mathcal{N}}\phi_i)\land(\land_{j\in\mathcal{M}}\phi^{\scriptscriptstyle O}_j)$ over the set $\Psi$. Then, we translate $\phi$ to a B\"uchi Automaton $\mathcal{BA}$ and 
we build the product $\widetilde{\mathcal{TS}} \coloneqq \mathcal{TS}\times\mathcal{BA}$. Using basic graph-search theory, we can find the accepting runs of $\widetilde{\mathcal{TS}}$ that satisfy $\phi$ and minimize the total cost $\chi$. These runs are directly projected to a sequence of desired states to be visited in the $\mathcal{TS}$. Although the semantics of LTL are defined over infinite sequences of services, it can be proven that there always exists a high-level plan that takes the form of a finite state sequence followed by an infinite repetition of another finite state sequence. For more details on the followed technique, the reader is referred to the related literature, e.g., \cite{baier2008principles}.

Following the aforementioned methodology, we obtain a high-level plan as sequences of states and atomic propositions $\pi_\text{pl} \coloneqq \pi_{s,1} \pi_{s,2}\dots$ and $\psi_\text{pl} \coloneqq \psi_{s,1} \psi_{s,1}\dots \models \phi$, which minimizes the cost $\chi$, with 
\begin{align*}
& \pi_{s,\ell} \coloneqq (\bar{\pi}_\ell,\bar{\pi}_{\scriptscriptstyle O,\ell},\bar{w}_\ell ) \in\Pi_s, \forall \ell\in\mathbb{N}, \notag \\
& \psi_{s,\ell} \coloneqq \Big(\bigcup\limits_{i\in\mathcal{N}}\psi_{i,\ell} \Big)\bigcup\Big(\bigcup\limits_{j\in\mathcal{M}}\psi^{\scriptscriptstyle O}_{j,\ell} \Big) \in 2^{\Psi}, \mathcal{L}(\pi_{s,\ell}), \forall \ell\in\mathbb{N},
\end{align*}
where 
\begin{itemize}
\item $\bar{\pi}_\ell \coloneqq \pi_{k_{1,\ell}}, \pi_{k_{2,\ell}}, \dots$ with $k_{i,\ell} \in\mathcal{K},\forall i\in\mathcal{N}$, 
\item $\bar{\pi}_{\scriptscriptstyle O,\ell} \coloneqq \pi_{k^{\scriptscriptstyle O}_{1,\ell}}, \pi_{k^{\scriptscriptstyle O}_{2,\ell}}, \dots$ with $k^{\scriptscriptstyle O}_{j,\ell} \in\mathcal{K},\forall j\in\mathcal{M}$, 
\item $\bar{w}_\ell \coloneqq w_{1,\ell}, w_{2,\ell},\dots$ with $w_{i,\ell} \in\mathcal{AG}_i, \forall i\in\mathcal{N}$,
\item $\psi_{i,\ell} \in 2^{\Psi_i}, \mathcal{L}_i(\pi_{k_{i,\ell}}), \forall i\in\mathcal{N}$, 
\item $\psi^{\scriptscriptstyle O}_{j,\ell}\in 2^{\Psi^{\scriptscriptstyle O}_j}, \mathcal{L}^{\scriptscriptstyle O}_j(\pi_{k^{\scriptscriptstyle O}_{j,\ell}}), \forall j\in\mathcal{M}$.
\end{itemize}
 
The path $\pi_\text{pl}$ is then projected to the individual sequences of the regions $\pi_{k^{\scriptscriptstyle O}_{j,1}} \pi_{k^{\scriptscriptstyle O}_{j,2}} \dots$ for each object $j\in\mathcal{M}$, as well as to the individual sequences of the regions $\pi_{k_{i,1}} \pi_{k_{i,2}} \dots$ and the boolean grasping variables $w_{i,1} w_{i,2}\dots$ for each agent $i\in\mathcal{N}$. The aforementioned sequences determine the behavior of agent $i\in\mathcal{N}$, i.e., the sequence of actions (transition, transportation, grasp, release or stay idle) it must take.

By the definition of $\mathcal{L}$ in Def. \ref{def:TS objects all agents}, we obtain that $\psi_{i,\ell} \in\mathcal{L}_i(\pi_{k_{i,\ell}}), \psi^{\scriptscriptstyle O}_{j,\ell} \in\mathcal{L}^{\scriptscriptstyle O}_j( \pi_{k^{\scriptscriptstyle O}_{j,\ell}}), \forall i\in\mathcal{N},j\in\mathcal{M}, \ell\in\mathbb{N}$. Therefore, since $\phi = (\land_{i\in\mathcal{N}}\phi_i)\land(\land_{j\in\mathcal{M}}\phi_{\scriptscriptstyle O_j})$ is satisfied by $\psi$, we conclude that $\psi_{i,1}\psi_{i,2}\dots \models \phi_i$ and $\psi^{\scriptscriptstyle O}_{j,1}\psi^{\scriptscriptstyle O}_{j,2}\dots\models \phi^{\scriptscriptstyle O}_j, \forall i\in\mathcal{N}, j\in\mathcal{M}$.


The sequences $\pi_{k_{i,1}}\pi_{k_{i,2}}\dots$, $\psi_{i,1}\psi_{i,2}\dots$ and $\pi_{k^{\scriptscriptstyle O}_{j,1}}\pi_{k^{\scriptscriptstyle O}_{j,2}}\dots, \psi^{\scriptscriptstyle O}_{j,1}\psi^{\scriptscriptstyle O}_{j,2}\dots$ over $\Pi, 2^{\Psi_i}$ and $\Pi, 2^{\Psi^{\scriptscriptstyle O}_j}$, respectively, produce the trajectories $\boldsymbol{q}_i(t)$ and $\boldsymbol{x}^{\scriptscriptstyle O}_j(t), \forall i\in\mathcal{N},j\in\mathcal{M}$. The corresponding behaviors are $\beta_i = (\boldsymbol{q}_i(t),\sigma_i) = (\boldsymbol{q}_i(t_{i,1}),\sigma_{i,1})(\boldsymbol{q}_i(t_{i,2}),\sigma_{i,2})\dots$ and $\beta^{\scriptscriptstyle O}_j$ $=$ $(\boldsymbol{x}^{\scriptscriptstyle O}_j(t),\sigma^{\scriptscriptstyle O}_j)= (\boldsymbol{x}^{\scriptscriptstyle O}_j(t^{\scriptscriptstyle O}_{j,1}),\sigma^{\scriptscriptstyle O}_{j,1})(\boldsymbol{x}^{\scriptscriptstyle O}_j(t^{\scriptscriptstyle O}_{j,2}),\sigma^{\scriptscriptstyle O}_{j,2})\dots$, respectively, according to Section \ref{subsec:PF}, with $\mathcal{A}_i(\boldsymbol{q}_i(t_{i,\ell}))\subset \pi_{k_{i,\ell}}, \sigma_{i,\ell}\in\mathcal{L}_i(\pi_{k_{i,\ell}})$ and $\mathcal{O}_j(\boldsymbol{x}_{\scriptscriptstyle O_j}(t_{\scriptscriptstyle O_{j,m}}))\in\pi_{k^{\scriptscriptstyle O}_{j,\ell}}, \sigma^{\scriptscriptstyle O}_{j,\ell} \in\mathcal{L}^{\scriptscriptstyle O}_j(\pi_{k^{\scriptscriptstyle O}_{j,\ell}})$. Thus, it is guaranteed that $\sigma_i \models \phi_i,\sigma^{\scriptscriptstyle O}_j \models \phi^{\scriptscriptstyle O}_j$ and consequently, the behaviors $\beta_i$ and $\beta^{\scriptscriptstyle O}_j$ satisfy the formulas $\phi_i$ and $\phi^{\scriptscriptstyle O}_j$, respectively, $\forall i\in\mathcal{N},j\in\mathcal{M}$. The aforementioned reasoning is summarized in the next theorem:
\begin{theorem}
The execution of the path $(\pi_{\text{pl}},\psi_{\text{pl}})$ of $\mathcal{TS}$ guarantees behaviors $\beta_i,\beta^{\scriptscriptstyle O}_j$ that yield the satisfaction of $\phi_i$ and $\phi^{\scriptscriptstyle O}_j$, respectively, $\forall i\in\mathcal{N},j\in\mathcal{M}$, providing, therefore, a solution to Problem \ref{problem}.  
\end{theorem}

\begin{remark}
Note that although the overall set of states of $\mathcal{TS}$ increases exponentially with respect to the number of agents/objects/regions, some states are not reachable, due to our constraints for the object transportation and the size of the regions, reducing thus the state complexity.
\end{remark}

\begin{figure}	
	\centering
	\includegraphics[scale=0.55,trim = 0cm 0cm 0cm 0cm]{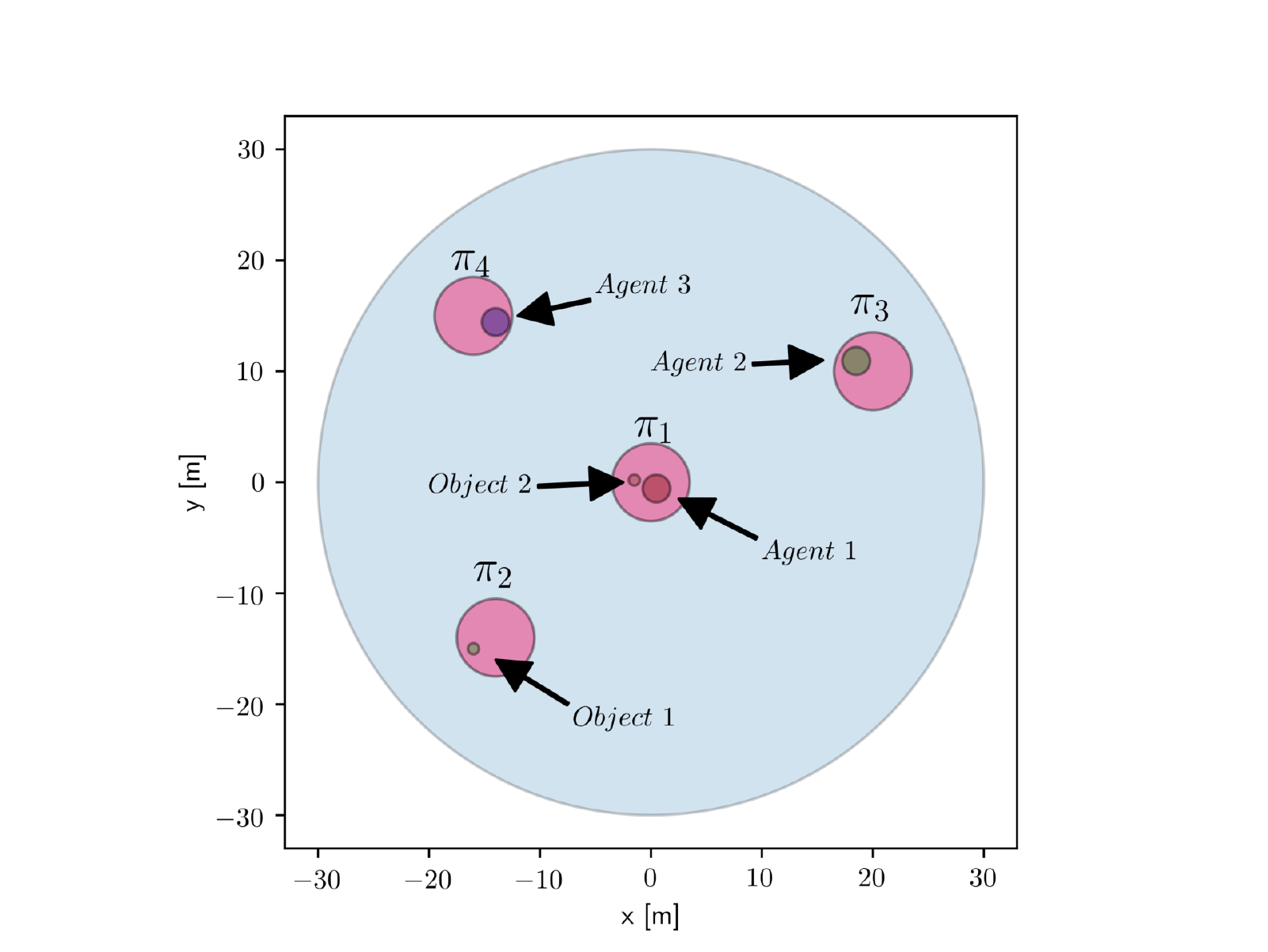}
	\caption{The initial workspace of the second simulation example, consisting of $3$ agents and $2$ objects. The agents and the objects are indicated via their corresponding radii. \label{fig:initial workspace}}
\end{figure}

\section{Simulation Results} \label{sec:simulations}

In this section we demonstrate our approach with computer simulations. We consider a workspace of radius $r_0 = 30 \text{m}$, with $K = 4$ regions of interest or radius $r_{\pi_k} = 3.5\text{m}$, $\forall k\in\mathcal{K}$, centered at $\boldsymbol{p}_{\pi_1} = (0,0,0), \boldsymbol{p}_{\pi_2} = (-14\text{m},-14\text{m},0), \boldsymbol{p}_{\pi_3} = (20\text{m},-10\text{m},0), \boldsymbol{p}_{\pi_4} = (-16 \text{m},15\text{m},0)$, respectively (see Fig. \ref{fig:initial workspace}). Moreover, we consider two cuboid objects of bounding radius $r^{\scriptscriptstyle O}_j = 0.5 \text{m}$, and mass $m^{\scriptscriptstyle O}_j = 0.5 \text{kg}$, $\forall j\in\{1,2\}$, initiated at  $\boldsymbol{x}^{\scriptscriptstyle O}_{1}(0) = [-16\text{m},15\text{m},0.5\text{m},0,0,0]^\top$ $\boldsymbol{x}^{\scriptscriptstyle O}_{2}(0) = [-1.5\text{m},0.2\text{m},0.5\text{m},0,0,0]^\top$, which implies that $\mathcal{O}_1(\boldsymbol{x}^{\scriptscriptstyle O}_1(0)) \subset \pi_2$, and $\mathcal{O}_2(\boldsymbol{x}^{\scriptscriptstyle O}_1(0)) \subset \pi_1$. The considered agents consist of a mobile base and a $2$-dof rotational robotic arm. The mobile base is rectangular with dimensions $0.5\text{m}\times 0.5\text{m}\times 0.2\text{m}$ and mass $0.5\text{kg}$, and the two arm links have length $1\text{m}$ and mass $0.5\text{kg}$ each.
	 The state vectors of the agents are $\boldsymbol{q}_i = [x_{c_i}, y_{c_i}, q_{i_1}, q_{i_2}]^\top\in\mathbb{R}^4, \dot{\boldsymbol{q}} = [\dot{x}_{c_i}, \dot{y}_{c_i}, \dot{q}_{i_1}, \dot{q}_{i_2}]^\top\in\mathbb{R}^4$, where $x_{c_i}, y_{c_i}$ are the planar position of the bases' center of mass, and $q_{i_1}, q_{i_2}$ the angles of the arms' joints. The geometric characteristics of the considered agents lead to a bounding radius of $r_i = 1.25 \text{m}$, $\forall i\in\mathcal{N}$. The atomic propositions are $\Psi_i =  \{ ``i\text{-}\pi_1",\dots,``i\text{-}\pi_4"\}$, $\forall i\in\mathcal{N}$, and $\Psi^{\scriptscriptstyle O} = \{ ``O_j\text{-}\pi_1",\dots,``O_j\text{-}\pi_4"\}$, $\forall j\in\mathcal{M}$, indicating whether the agents/objects are in the corresponding regions. The labeling functions are, therefore, $\mathcal{L}_i(\pi_k) = \{ ``i\text{-}\pi_k" \}$, $\mathcal{L}^{\scriptscriptstyle O}_j(\pi_k) = \{ ``O_j\text{-}\pi_k" \}$, $\forall k\in\mathcal{K},i\in\mathcal{N},j\in\mathcal{M}$.
We test two scenarios with $N=2, 3$ agents, respectively. We generate the optimal high-level plan for these scenarios and present two indicative transitions of the continuous execution for the second case. The simulations were carried out using Python environment on a laptop computer with $4$ cores at $2.6$GHz CPU and $8$GB of RAM memory.

\begin{table}[t]
	\caption{The agent actions for the discrete path of the first simulation example}
	\label{table:Path 1}
	\centering
	\begin{tabular}{||c| c| c |c | ||} 
		\hline
		 $\pi_{s,\ell} $ & Actions & $\pi_{s,\ell} $ & Actions \\ [0.5ex] 
		\hline\hline 		
		$\pi_{s,1}$ & ($-$) & $\pi_{s,14}$ & ($\pi_1 \xrightarrow{T}_{\{1,2\},2} \pi_2) $  \\ [0.5ex] 
		\hline
		$\pi_{s,2}$ & ($-$, $\pi_3 \to_2 \pi_1$)  & $\pi_{s,15}$ & ($1 \xrightarrow{r} 2$, $2 \xrightarrow{r} 2$)   \\ [0.5ex] 
		\hline
		$\pi_{s,3}$ & ($1 \xrightarrow{g} 2$, $2 \xrightarrow{g} 2$)   & $\pi_{s,16}$ & ($\pi_2 \to_2 \pi_4$, $\pi_2 \to_2 \pi_4$)   \\ [0.5ex]
		\hline
		$\pi_{s,4}$ &  ($\pi_1 \xrightarrow{T}_{\{1,2\},2} \pi_4 $)  &  $\pi_{s,17}$ & ($1 \xrightarrow{g} 1$, $2 \xrightarrow{g} 1$)  \\ [0.5ex]
		\hline
		$\pi_{s,5}$ &  ($\pi_4 \xrightarrow{T}_{\{1,2\},2} \pi_1 $)  & $\pi_{s,18}$ & ($\pi_4 \xrightarrow{T}_{\{1,2\},1} \pi_1 $)  \\ [0.5ex]
		\hline
		$\pi_{s,6}$ & ($1 \xrightarrow{r} 2$, $2 \xrightarrow{r} 2$)  & $\pi_{s,19}$ & ($\pi_1 \xrightarrow{T}_{\{1,2\},1} \pi_4 $)  \\ [0.5ex]
		\hline
		$\pi_{s,7}$ & ($\pi_1 \to_1 \pi_2$, $\pi_1 \to_2 \pi_2$)   & $\pi^\star_{s,20}$ & ($-$, $2 \xrightarrow{r} 1$)   \\ [0.5ex]
		\hline 
		$\pi_{s,8}$ & ($1 \xrightarrow{g} 1$, $2 \xrightarrow{g} 1$)  & $\pi^\star_{s,21}$ & ($-$, $\pi_4 \to_2 \pi_3$)   \\ [0.5ex]
		\hline 
		$\pi_{s,9}$ & ($\pi_2 \xrightarrow{T}_{\{1,2\},1} \pi_4 $)  & $\pi^\star_{s,22}$ & ($-$, $\pi_3 \to_2 \pi_4$)   \\ [0.5ex]
		\hline
		$\pi_{s,10}$ & ($1 \xrightarrow{r} 1$, $2 \xrightarrow{r} 1$)  & $\pi^\star_{s,23}$ & ($-$, $2 \xrightarrow{g} 1$)   \\ [0.5ex]
		\hline 
		$\pi_{s,11}$ & ($-$, $\pi_4 \to_2 \pi_3$)  & $\pi^\star_{s,24}$ & ($\pi_4 \xrightarrow{T}_{\{1,2\},1} \pi_1 $)  \\ [0.5ex]
		\hline
		$\pi_{s,12}$ & ($\pi_4 \to_2 \pi_1$, $\pi_3 \to_2 \pi_1$)  & 	$\pi^\star_{s,25}$ & ($\pi_1 \xrightarrow{T}_{\{1,2\},1} \pi_4 $)   \\ [0.5ex]	
		\hline 
		$\pi_{s,13}$ & ($1 \xrightarrow{g} 2$, $2 \xrightarrow{g} 2$) & &  \\ [0.5ex]
		\hline
	\end{tabular}
\end{table}

\begin{enumerate}[label= \underline{\textbf{Case \roman*}},align=left, leftmargin=0pt, listparindent=\parindent, labelwidth=0pt, itemindent=!]
	\item:  We consider $N=2$ agents with initial conditions $\boldsymbol{q}_1(0) = [0.5\text{m},0,\frac{\pi}{4}\text{rad},\frac{\pi}{4}\text{rad}]^\top,  \boldsymbol{q}_2(0) = [18.5\text{m},11.5\text{m},\frac{\pi}{4}\text{rad},\frac{\pi}{4}\text{rad}]^\top$, $\dot{\boldsymbol{q}}_i(0) = [0,0,0,0]^\top, \forall i\in\{1,2\}$ which imply that $\mathcal{A}_1(\boldsymbol{q}_1(0)) \subset \pi_1$, $\mathcal{A}_2(\boldsymbol{q}_2(0)) \subset \pi_3$, and that no collisions occur at $t=0$, i.e., $\mathcal{C}_{1,2}(\boldsymbol{q}_1(0), \boldsymbol{q}_2(0)) = \mathcal{C}_{\scriptscriptstyle O_1, \scriptscriptstyle O_2}(\boldsymbol{x}^{\scriptscriptstyle O}_1(0), \boldsymbol{x}^{\scriptscriptstyle O}_2(0)) = \mathcal{C}_{i,\scriptscriptstyle O_j}(\boldsymbol{q}_1(0), \boldsymbol{x}^{\scriptscriptstyle O}_j(0)) = \bot, \forall (i,j)\in\{1,2\}\times\{1,2\}$. 
	 We also assume that $\mathcal{AG}_{i,0}(\boldsymbol{q}_i(0),\boldsymbol{x}^{\scriptscriptstyle O}(0)) = \top, \forall i\in\{1,2\}$.  We represent the agents' power capabilities with the scalars $\zeta_1 = 2, \zeta_2 = 4$ and construct the functions $\Lambda(m^{\scriptscriptstyle O}_1, \zeta_\mathcal{T}) = \top $ if and only if $\sum_{\tau\in\mathcal{T}} \zeta_\tau \geq 5$, with $\mathcal{AG}_{\tau,1} = \top \Leftrightarrow \tau\in\mathcal{T}$, and $\Lambda(m^{\scriptscriptstyle O}_2, \zeta_\mathcal{T}) = \top$ if and only if $\sum_{\tau\in\mathcal{T}} \zeta_\tau \geq 6$, with $\mathcal{AG}_{\tau,2} = \top \Leftrightarrow \tau\in\mathcal{T}$, i.e., the objects can be transported only if the agents that grasp them have a sum of capability scalars no less than $5$ and $6$, respectively. Regarding the cost $\chi$, we simply choose the sum of the distances of the transition and transportation regions, i.e., given $\pi_s, \widetilde{\pi}_s$ as in \eqref{eq:pi_s} such that $\pi_s \to_s \widetilde{\pi}_s$, we have that $\chi = \sum_{i\in\{1,2\}} \{ \|\boldsymbol{p}_{\pi_{k_i}} - \boldsymbol{p}_{\pi_{\widetilde{k}_i}} \|^2 \} + \sum_{j\in\{1,2\}}\|\boldsymbol{p}_{\pi_{k^{\scriptscriptstyle O}_j}} - \boldsymbol{p}_{\pi_{\widetilde{k}^{\scriptscriptstyle O}_j}} \|^2 \} $. The LTL formula is taken as $ (\square \neg ``1\text{-}\pi_3" ) \land ( \square\lozenge ``2\text{-}\pi_3" ) \land ( \square \lozenge ``O_1\text{-}\pi_1" )\land \square(``O_1\text{-}\pi_1" \to \bigcirc ``O_1\text{-}\pi_4") \land (\lozenge ``O_2\text{-}\pi_4")$, which represents the following behavior. Agent $1$ must never go to region $\pi_3$, which must be visited by agent $2$ infinitely many times, object $1$ must be taken infinitely often to region $\pi_1$, always followed by a visit in region $\pi_4$, and object $2$ must be eventually taken to region $\pi_4$.
	 
	 The resulting transition system $\mathcal{TS}$ consists of $560$ reachable states and $7680$ transitions and it was created in $3.19 \sec$. The B\"uchi automaton $\mathcal{BA}$ contains $7$ states and $29$ transitions  and the product $\widetilde{\mathcal{TS}}$ contains $3920$ states and $50976$ transitions. Table \ref{table:Path 1} shows the actions of the agents for the derived path, which is the sequence of states $\pi_{s,1}\pi_{s,2}\dots...(\pi^\star_{s,20},\dots,\pi^\star_{s,25})^\omega$, where the states with ($^\star$) constitute the suffix that is run infinitely many times.  
	 Loosely speaking, the derived path describes the following behavior: Agent $2$ goes first to $\pi_1$ to grasp and transfer object $2$ to $\pi_4$ and back to $\pi_1$ with agent $1$. The two agents then navigate to $\pi_2$ to take object $1$ to $\pi_4$. In the following, after agent $2$ goes to $\pi_3$,  they both go to $\pi_1$ to transfer object $2$ to $\pi_2$. Then, they navigate to $\pi_4$ to transfer object $1$ to $\pi_1$ and back.  Finally, the actions that are run infinitely many times consist of agent $2$ going to from $\pi_4$ to $\pi_3$ and back, and transferring object $1$ to $\pi_1$ and $\pi_4$ with agent $1$.  
    	One can verify that the resulting path satisfies the LTL formula. Note also that the regions are not large enough to contain both agents and objects in a grasping configuration, which played an important role in the derivation of the plan.
	 	The time taken for the construction of the product $\widetilde{\mathcal{TS}}$ and the derivation of the path was $2.79 \sec$.

	 \item We now consider $N=3$ agents with $\boldsymbol{q}_1(0)$, $\boldsymbol{q}_2(0)$ as in case (i), $\boldsymbol{q}_3(0) = [-14\text{m}, 15\text{m},\frac{\pi}{4}\text{rad}, \frac{\pi}{4}\text{rad}]^\top \implies \mathcal{A}_3(\boldsymbol{q}_3(0))\in \pi_4$, $\mathcal{AG}_{3,0}(\boldsymbol{q}_i(0),\boldsymbol{x}^{\scriptscriptstyle O}(0)) = \top$, $\zeta_3 = 3$, and no collisions occurring at $t=0$. The functions $\Lambda$ and $\chi$ are the same as in case (i). The formula in this scenario is $(\square \neg ``1\text{-}\pi_3")\land (\square\lozenge ``2\text{-}\pi_3") \land (\square\lozenge ``O_1\text{-}\pi_1") \land \square(``O_1\text{-}\pi_1" \to \lozenge  ``O_1\text{-}\pi_4") \land (\square\lozenge ``O_2\text{-}\pi_3")$, which represents the following behavior. Agent $1$ must never visit region $\pi_3$, which must be visited infinitely many times by agent $2$, object $1$ must be taken infinitely many times to region $\pi_1$, eventually followed by a visit in region $\pi_4$, and object $2$ must be taken infinitely many times to region $\pi_2$.
	 
	 
	 The resulting transition system $\mathcal{TS}$ consists of $3112$ reachable states and $154960$ transitions and it was created in $100.74 \sec$. The B\"uchi automaton $\mathcal{BA}$ contains $9$ states and $49$ transitions  and the product $\widetilde{\mathcal{TS}}$ contains $28008$ states and $1890625$ transitions. Table \ref{table:Path 2 actions} shows the agent actions for the derived path as the sequence of states $\pi_{s,1}\pi_{s,2}\dots...(\pi^\star_{s,10},\pi^\star_{s,11})^\omega$. 
	 In this case, the three agents navigate first to regions $\pi_2,\pi_1$, and $\pi_1$, respectively, and agents $2$ and $3$ take object $2$ to $\pi_3$. Next, agent $3$ goes to $\pi_2$ to transfer object $1$ to $\pi_1$ and then $\pi_4$ with agent $1$. The latter transportations occur infinitely often.	 	 	
	 	The time taken for the construction of the product $\widetilde{\mathcal{TS}}$ and the derivation of the path was $4573.89 \sec$. It is worth noting the exponential increase of the computation time with the simple addition of just one agent, which can be attributed to the centralized manner of the proposed methodology. The necessity, therefore, of less computational, decentralized schemes is evident and constitutes the main focus of our future directions.
\end{enumerate}

\begin{figure}	
	\centering
	\includegraphics[scale=0.55,trim = 0cm 0cm 0cm 0cm]{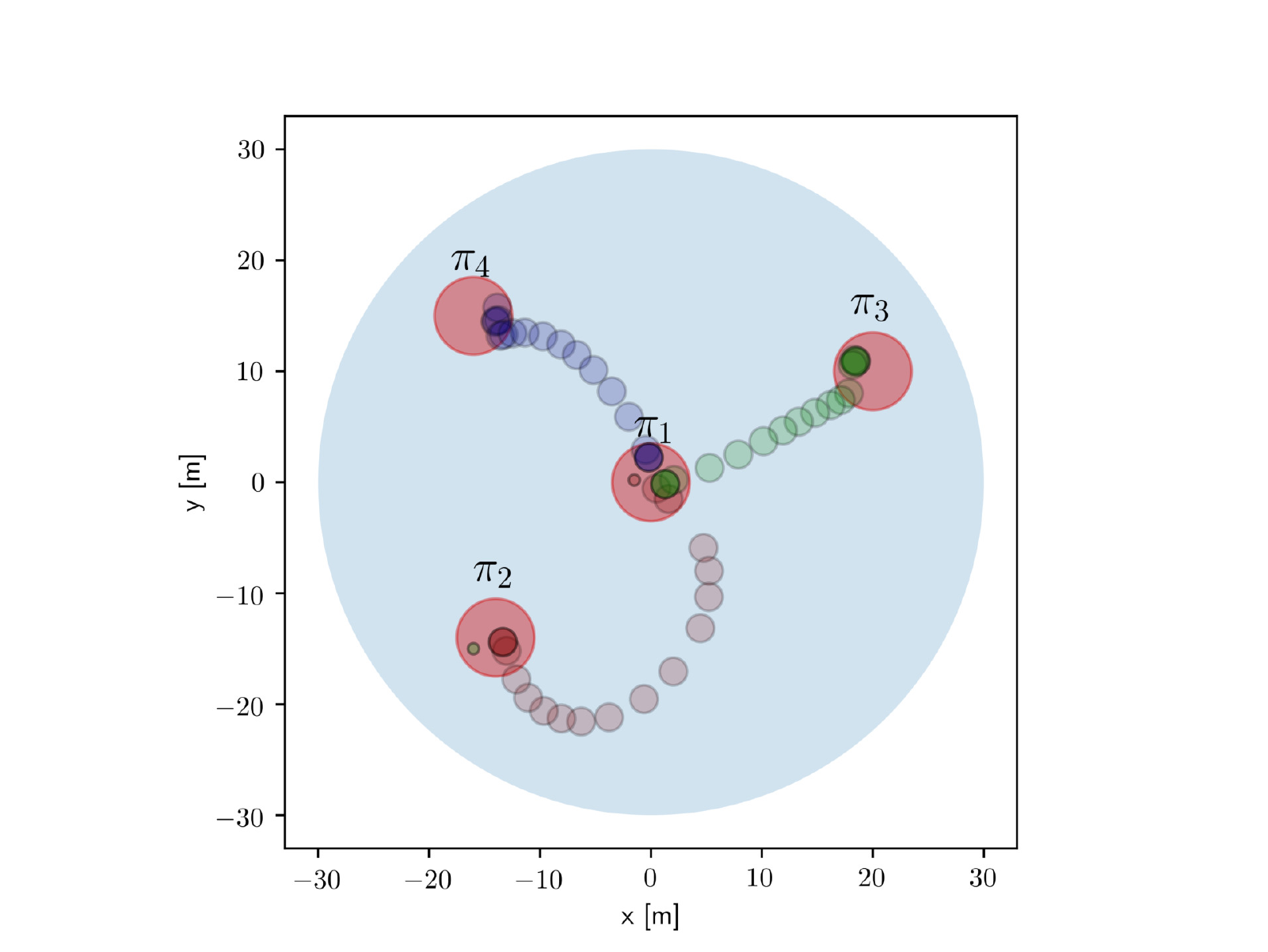}
	\caption{The transition $\pi_{s,1}\to_s \pi_{s,2}$ (a), that corresponds to the navigation of the agents $\pi_1\to_1\pi_2$, $\pi_3\to_2\pi_1$, $\pi_4\to_3\pi_1$. \label{fig:transition_1}}
\end{figure}

\begin{figure}	
	\centering
	\includegraphics[scale=0.55,trim = 0cm 0cm 0cm 0cm]{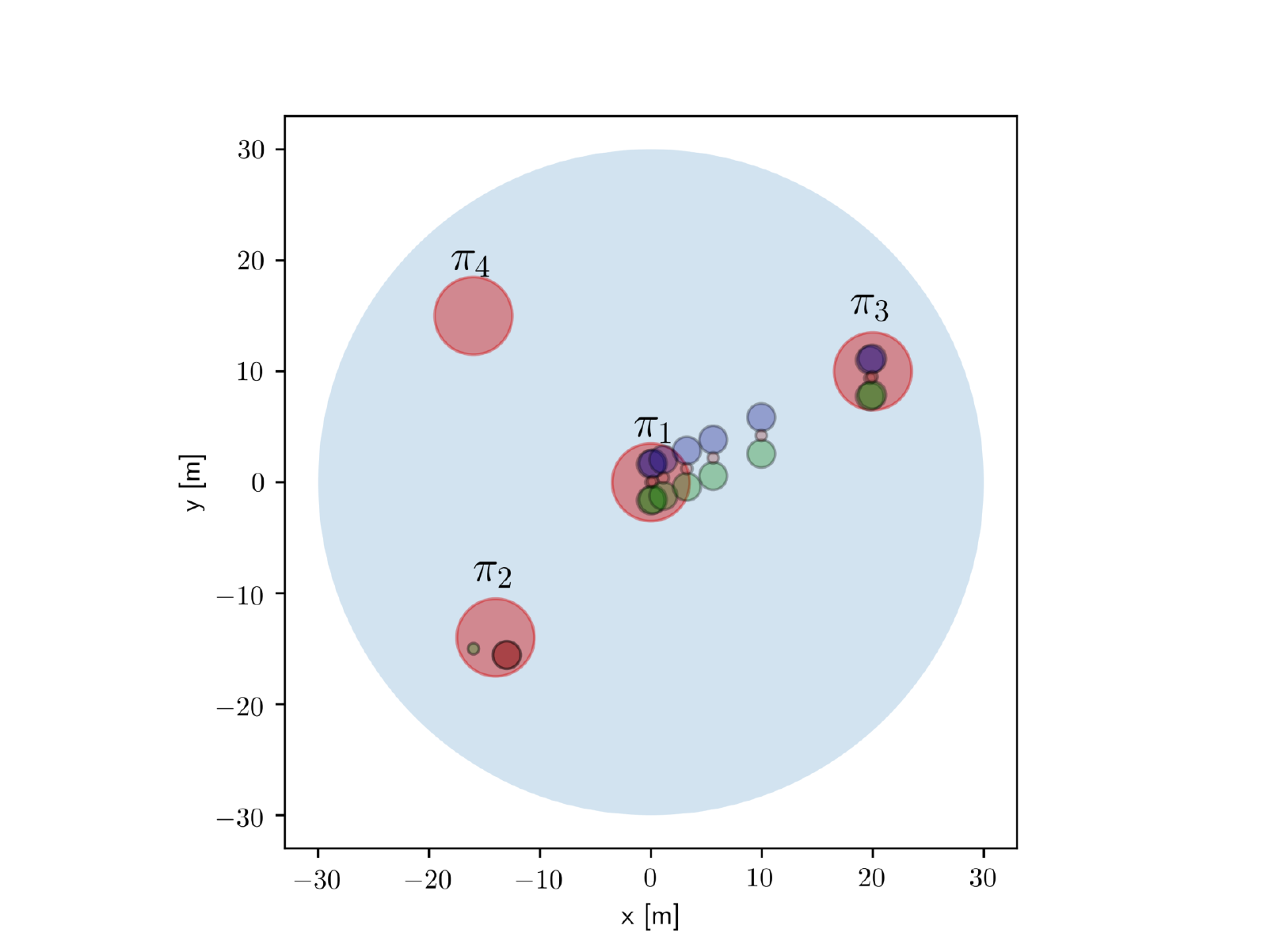}
	\caption{The transition $\pi_{s,3}\to_s \pi_{s,4}$ (b), that corresponds to the transportation $\pi_1 \xrightarrow{T}_{\{2,3\}} \pi_3$. \label{fig:transition_2}}
\end{figure}

Next, we present the continuous execution of the transitions $\pi_{s,1} \to_s \pi_{s,2}$, and $\pi_{s,3}\to_s\pi_{s,4}$ for the second simulation scenario. More specifically, Fig. \ref{fig:transition_1} depicts the navigation of the three agents $\pi_1\to_1\pi_2$, $\pi_3\to_2\pi_1$, and $\pi_4\to_3\pi_1$, that corresponds to $\pi_{s,1}\to_s \pi_{s,2}$, with gains $\boldsymbol{K}_z = \text{diag}\{0.01,0.01,0.01\}$, $\forall z\in\{1,2,3\}$, and which had a duration of $900 \sec$. Moreover, Fig. \ref{fig:transition_2} depicts the transportation of object $2$ by agents $2$ and $3$, i.e., $\pi_1 \xrightarrow{T}_{\{2,3\}} \pi_3$, that corresponds to $\pi_{s,3}\to_s\pi_{s,4}$, with load sharing coefficients $c_1 = c_2 = 0.5$, and corresponding time duration $300 \sec$.

\begin{table}[t]
	\caption{The agent actions for the discrete path of the second simulation example}
	\label{table:Path 2 actions}
	\centering
	\begin{tabular}{||c| c ||} 
		\hline
		$\pi_{s,\ell} $ &  Actions\\ [0.5ex] 
		\hline\hline 		
		$\pi_{s,1}$ & ($-$) \\ [0.5ex] 	
		\hline 
		$\pi_{s,2}$ & ($\pi_1 \to_1 \pi_2, \pi_3 \to_2 \pi_1 , \pi_4\to_3\pi_1$) \\ [0.5ex] 	
		\hline 
		$\pi_{s,3}$ & ($-, 2\xrightarrow{g}1, 3\xrightarrow{g}2 $) \\ [0.5ex] 
		\hline 
		$\pi_{s,4}$ & ($-,\pi_1 \xrightarrow{T}_{\{2,3\},2} \pi_3,$) \\ [0.5ex] 
		\hline 
		$\pi_{s,5}$ & ($-,-, 3\xrightarrow{r}2$) \\ [0.5ex]
		\hline 
		$\pi_{s,6}$ & ($-,-,\pi_3\to_3\pi_2$) \\ [0.5ex]
		\hline 
		$\pi_{s,7}$ & ($1\xrightarrow{g}1, 3\xrightarrow{g}1$) \\ [0.5ex]
		\hline 
		$\pi_{s,8}$ & ($\pi_2 \xrightarrow{T}_{\{1,3\},1} \pi_1,-$) \\ [0.5ex] 
		\hline 
		$\pi_{s,9}$ & ($\pi_1 \xrightarrow{T}_{\{1,3\},1} \pi_4,-$) \\ [0.5ex] 
		\hline 
		$\pi^\star_{s,10}$ & ($\pi_4 \xrightarrow{T}_{\{1,3\},1} \pi_1,-$) \\ [0.5ex] 
		\hline 
		$\pi^\star_{s,11}$ & ($\pi_1 \xrightarrow{T}_{\{1,3\},1} \pi_4,-$) \\ [0.5ex] 
		\hline
	\end{tabular}
\end{table}

\section{Conclusion} \label{sec:conclusion}
We have presented a novel hybrid control framework for the motion planning of a system comprising of $N$ agents and $M$ objects. We designed appropriate continuous control protocols that guarantee the agent transition and object transportation among predefined regions of interest. In that way, the coupled multi-agent system is abstracted in a finite transition system, which is used to derive plans that satisfy complex LTL formulas.  Future works will address decentralization of the framework by incorporating limited sensing information for the agents, as well as real-time experiments.


%

\ifCLASSOPTIONcaptionsoff
  \newpage
\fi



%

\bibliographystyle{IEEEtran}
\bibliography{references}  
\end{document}